\pdfoutput=1
\documentclass[twocolumn]{aastex63}
\usepackage{natbib}
\usepackage{hyperref}
\usepackage{amsmath}
\usepackage{float}
\usepackage{threeparttable}
\usepackage{lineno}

\newcommand{\Msolar}{$M_{\odot}$}

\newcommand{\micro}{\,$\mu$m}
\newcommand{\simi}{$\sim$}
\newcommand{\as}{$''$}

\newcommand{\kps}{km s$^{-1}$}

\newcommand{\htwo}{H\,{\sc ii}}

\newcommand{\cone}{Class\,{\sc I} YSOs}

\newcommand{\cetno}{C$^{18}$O($J$ = 2--1)}
\newcommand{\tco}{$^{13}$CO($J$ = 2--1)}
\shorttitle{G321.93$-$0.01: A Rare Site of Multiple HFSs}
\shortauthors{Maity et al.}

\begin{document}
\title{G321.93$-$0.01: A Rare Site of Multiple Hub-Filament Systems with Evidence of Collision and Merging of Filaments}
\correspondingauthor{A.~K.  Maity}
\email{Email: aruokumarmaity123@gmail.com}
\author[0000-0002-7367-9355]{A.~K. Maity}
\affiliation{Astronomy \& Astrophysics Division, Physical Research Laboratory, Navrangpura, Ahmedabad 380009, India}
\affiliation{Indian Institute of Technology Gandhinagar Palaj, Gandhinagar 382355, India}

\author[0000-0001-6725-0483]{L.~K. Dewangan}
\affiliation{Astronomy \& Astrophysics Division, Physical Research Laboratory, Navrangpura, Ahmedabad 380009, India}

\author[0000-0001-8812-8460]{N.~K.~Bhadari}
\affiliation{Astronomy \& Astrophysics Division, Physical Research Laboratory, Navrangpura, Ahmedabad 380009, India}

\author[0000-0002-8966-9856]{Y.~Fukui}
\affiliation{Department of Physics, Nagoya University, Furo-cho, Chikusa-ku, Nagoya 464-8601, Japan}

\author[0000-0003-4941-5154]{A. {Haj Ismail}}
\affiliation{College of Humanities and Sciences, Ajman University, 346 Ajman, United Arab Emirates}

\author[0009-0001-2896-1896]{O.~R.~Jadhav}
\affiliation{Astronomy \& Astrophysics Division, Physical Research Laboratory, Navrangpura, Ahmedabad 380009, India}
\affiliation{Indian Institute of Technology Gandhinagar Palaj, Gandhinagar 382355, India}

\author[0000-0001-5731-3057]{Saurabh Sharma}
\affiliation{Aryabhatta Research Institute of Observational Sciences, Manora Peak, Nainital 263001, India}

\author[0000-0003-2062-5692]{H.~Sano}
\affiliation{Faculty of Engineering, Gifu University, 1-1 Yanagido, Gifu 501-1193, Japan}

\begin{abstract}
Hub-filament systems (HFSs) are potential sites of massive star formation (MSF). To understand the role of filaments in MSF and the origin of HFSs, we conducted a multi-scale and multi-wavelength observational investigation of the molecular cloud G321.93--0.01. The {\tco} data reveal multiple HFSs, namely, HFS-1, HFS-2, and a candidate HFS (C-HFS). HFS-1 and HFS-2 exhibit significant mass accretion rates ($\dot{M}_{||}$ $> 10^{-3}$ $M_{\odot}$ yr$^{-1}$) to their hubs (i.e., Hub-1 and Hub-2, respectively). Hub-1 is comparatively massive, having higher $\dot{M}_{||}$ than Hub-2, allowing to derive a relationship $\dot{M}_{||} \propto M^{\beta}_{\rm{hub}}$, with $\beta \sim1.28$. Detection of three compact {\htwo} regions within Hub-1 using MeerKAT 1.28 GHz radio continuum data and the presence of a clump (ATL-3), which meets Kauffmann \& Pillai's criteria for MSF, confirm the massive star-forming activity in HFS-1. We find several low-mass ALMA cores (1--9\,{\Msolar}) inside ATL-3. The presence of a compact {\htwo} region at the hub of C-HFS confirms that it is active in MSF. Therefore, HFS-1 and C-HFS are in relatively evolved stages of MSF, where massive stars have begun ionizing their surroundings. Conversely, despite a high $\dot{M}_{||}$, the non-detection of radio continuum emission toward Hub-2 suggests it is in the relatively early stages of MSF. Analysis of {\tco} data reveals that the formation of HFS-1 was likely triggered by the collision of a filamentary cloud about 1 Myr ago. In contrast, the relative velocities ($\gtrsim 1$ {\kps}) among the filaments of HFS-2 and C-HFS indicate their formation through the merging of filaments.

\end{abstract}
\section{Introduction}
\label{sec_intro_G322}
Since last two decades, hub-filament systems (HFSs) have emerged as the potential sites of massive star
formation \citep[MSF; e.g.,][and references therein]{myers_2009,Kumar_2020,Anderson_2021MNRAS,Bhadari_2022ApJ,dewangan_hfs_2017,Dewangan_2024MNRAS}. 
From the cloud scale, molecular gas and dust get collected through the filaments to the hub, as a result, hub becomes the densest region of the system with a typical H$_2$ column density, $N$(H$_2$) $>$ 10$^{22}$ cm$^{-2}$ \citep[e.g.,][]{trevino19,Wang_2019,dewangan_2018_dec,Dewangan_2023JApA,Seshadri_2024MNRAS}. The higher column density at the hub leads to a high mass accretion rate onto the (proto-)stars and, hence, the formation of massive stars \citep[][]{McKee_2003ApJ,zinnecker07,motte_2018}.
A recent study by \citet{Kumar_2020} found a large number (nearly 3500) of HFSs in the Galactic plane. They also discovered that all massive stars above 100 {\Msolar} form inside the hubs. Although the potential of HFSs in the context of MSF is well recognized, their origin remains elusive. In this context, ideas such as overlapping or merging of gravitationally unbound flow-driven filaments \citep{Kumar_2020} and collision of molecular clouds \cite[][]{Balfour_2017MNRAS,Maity_2024arXiv} are put forward in recent years to explain the origin of HFSs.
The scenario of overlapping filaments effectively explains the origin of the HFS at G083.097+03.270 \citep{Panja_2023ApJ}. However, reconciling the complex morphologies of HFSs observed in studies, such as Monoceros R2 \citep{trevino19}, G45.3+0.1 \citep{Bhadari_2022ApJ},  N159E-Papillon \citep{fukui19ex} and N159W-South \citep{tokuda19ex} with the model of \citet{Kumar_2020} can be challenging. According to \citet{Maity_2024arXiv}, the formation of HFSs from cloud-cloud collision (CCC) is a combined effect of turbulence, shock compression, magnetic fields, and gravity. The connection between CCC and the presence of HFSs has already been observed in several Galactic \citep[e.g., SDC13, W31 complex, W33 complex, and AFGL 5180 \& 6366S;][]{Peretto_2014,Wang_2022ApJ,maity_W31,Zhou_2023MNRAS,Maity_2023MNRAS} and extragalactic \citep[e.g., N159E-Papillon and N159W-South;][]{fukui19ex,tokuda19ex} massive star-forming regions (MSFRs).
However, each molecular cloud has a distinct gas distribution in both position and velocity space. Therefore, detailed observational studies of new target sites are essential to enhance our understanding of the proposed ideas of HFS formation and, consequently, to obtain comprehensive insights into the subject. The observational investigations are also essential to determine how the mass accretion rate through the filaments affects overall star-forming activity, including MSF.

Massive stars are known for their tremendous radiative and mechanical feedback, which greatly influences their parental molecular environment \citep[][]{Stromgren_1939ApJ,Bressert_2012_new,Dale_2014MNRAS,Geen_2015MNRAS}. Therefore, the study of the origin of HFSs and the effect of mass accretion through the filaments on star-forming activity demands comprehensive investigations at the early stages of MSF, which ensures an undisturbed molecular environment. MSF begins inside hot molecular cores \citep[][]{Mayra_1999,Tak_2004,Paron_2024}. Later, the intense ultraviolet radiation from massive stars, beyond the Lyman limit, ionizes the surrounding gas \citep[][]{Panagia_1973}. The compactness of the ionized ({\htwo}) regions serves as a proxy to detect the early stages of MSF \citep[see][and references therein]{Maity_2023MNRAS}.

This work involves a multi-scale and multi-wavelength observational investigation of the molecular cloud G321.8$-$0.01 (hereafter, G321), which is at a distance ($d$) of about 1.98 kpc \citep{Atlasgal}. Figure~\ref{G321_herschel_maps}a displays {\it Herschel} three-color composite image (red: 500\,$\mu$m, green: 350\,$\mu$m, and blue: 250\,$\mu$m) of our target site. Several parsec-scale filaments appear in the visual inspection of the color composite map. The map is overlaid with the Sydney University Molonglo Sky Survey \citep[SUMSS;][]{Bock_1999AJ} 843 MHz radio continuum emission contours and the position of the APEX Telescope Large Area Survey of the Galaxy (ATLASGAL) clumps \citep{Atlasgal}. Earlier studies toward G321 detected 22 GHz H$_2$O and Class~I 95 GHz CH$_3$OH maser emissions \citep{Walsh_2011MNRAS,Chen_2011ApJ,Yang_2017ApJS}, marked with the plus symbol in Figure~\ref{G321_herschel_maps}a. Additionally, \citet{Yang2022} detected outflow signatures in several ATLASGAL clumps, represented by black asterisks in Figure~\ref{G321_herschel_maps}a, which are indications of ongoing star formation in G321.  
The presence of filaments in the {\it Herschel} continuum images makes G321 an ideal site to search for new HFSs and to investigate their possible formation mechanisms. The absence of any extended {\htwo} region ensures that the gas kinematics remain unaffected by feedback from massive stars. This allows us to explore the initial conditions of star and/or structure formation within the parental molecular cloud. Therefore, such source offers a unique opportunity to study the early stages of HFSs, which is a challenging task in star formation research. The proximity of G321 ($d$ {\simi}1.98 kpc) enables the detection and study of filaments using the SEDIGISM molecular line data. This study helps to determine the mass accretion rate through the filaments and understand their impact on star-forming activity, including MSF.

The structure of this paper is as follows: Section~\ref{sec_obser_G322} provides detailed descriptions of the dataset used in our research. Section~\ref{sec_results_G321} outlines the major findings. These findings are analyzed and interpreted in Section~\ref{G321_Discussion}. This paper concludes with Section~\ref{sec_sum_JP}, highlighting the principal conclusions drawn from the results of this study.
\section{Data sets}
\label{sec_obser_G322}
In this work, we explored several archival data sets from different surveys spanning in wavelengths from near-infrared (NIR) to radio, which are listed in Table~\ref{tab_data_g321}. The ionized emission toward our target region is traced with SUMSS 843 MHz \citep{Bock_1999AJ} and SARAO MeerKAT Galactic Plane Survey (SMGPS) 1.28 GHz \citep{Goedhart_2024MNRAS} radio continuum images. The rms noise levels ($\sigma$) of the SUMSS and SMGPS data are about 1 mJy beam$^{-1}$ and 20 $\mu$Jy beam$^{-1}$, respectively \citep{Bock_1999AJ,Goedhart_2024MNRAS}.
%
We used $^{13}$CO($J$ = 2--1) and C$^{18}$O($J$ = 2--1) line data from the Structure, Excitation, and Dynamics of the Inner Galactic Interstellar Medium \citep[SEDIGISM;][]{schuller17} survey to study the gas distribution and kinematics, and to calculate the mass of the molecular gas. 
The native resolution, velocity separation, and rms noise of these line data are approximately 30$''$, 0.25 {\kps} and 1 K, respectively \citep[see][for more details]{schuller17}. To reduce noise and enhance the visibility of faint or diffuse features in the molecular line data without significantly distorting the overall structure, the data were smoothed using a Gaussian function with a width of 2 pixels (i.e., {\simi}19$''$). The resultant angular resolution and rms noise are about 36$''$ and 0.35 K, respectively. 
%
The {\it Herschel} images at 160--500 $\mu$m and the ATLASGAL 870 $\mu$m image are examined for the large-scale ({\simi}1-10\,pc) features in the thermal dust emission. In addition, these {\it Herschel} images are also utilized to produce the column density (i.e., $N$(H$_2$)) and dust temperature ($T_\mathrm{d}$) maps (see Section~\ref{sec_phy_env_G321} for details). 
For the core-scale study ({\simi}0.01\,pc), high-resolution Band-7 ({\simi}0.88 mm) images from Atacama Large Millimeter/submillimeter Array (ALMA; Proposal Id: 2013.1.00960.S, PI: Csengeri, Timea) at two different resolutions were collected from the ALMA Archive{\footnote{https://jvo.nao.ac.jp/portal/alma/archive.do}} (see Table~\ref{tab_data_g321}). 
The ALMA observations in two different configurations, the 7 m and 12 m arrays, provided images with different resolutions. The 12 m array provides higher spatial resolution, while 7 m array offer better sensitivity for extended sources. Therefore, the 7 m array and 12 m array together enable detailed studies of both the diffuse and compact components of the target.
%
The photometric magnitudes of point-like sources at 3.6, 4.5, and 5.8 $\mu$m were acquired from {\it Spitzer} GLIMPSE-I Spring $' $07 catalog. Additionally, we collected Vista Variables in the V{\'i}a L{\'i}actea (VVV) $K_{\mathrm s}$ band image and photometric magnitudes of point-like sources in the $H$ and $K_{\mathrm s}$ bands from the VVV Data Release 5 \citep[DR5;][]{McMahon_2021yCat}. 
For bright sources that are saturated in the VVV survey, we used photometric data from Two Micron All Sky Survey \citep[2MASS;][]{Minniti_2010, Minniti_2017yCat}.
\begin{table*}
\scriptsize
\centering
\caption{List of archive data sets utilized in this work.}
\label{tab1}
\begin{tabular}{lcccl}
\hline \hline
  Survey  &  Wavelength/Frequency      &  Resolution (arcsec)        &  Reference \\
    &  /line(s)       &  /Beam size (arcsec$^{2}$)        &   \\
\hline
SARAO MeerKAT Galactic Plane Survey (SMGPS) & 1.28 GHz  & $\sim$8$\times$8& \citet{Goedhart_2024MNRAS}\\
Sydney University Molonglo Sky Survey (SUMSS) & 843 MHz  & $\sim$45$\times$63& \citet{Bock_1999AJ}\\
SEDIGISM &  $^{13}$CO/C$^{18}$O ($J$ = 2--1) & $\sim$30 &\citet{schuller17}\\
Atacama Large Millimeter/submillimeter Array (ALMA)  &879 $\mu$m & $\sim$4.9$\times$3.1 and 0.33$\times$0.16  &  Pro. Id: 2013.1.00960.S \\
ATLASGAL  &870 $\mu$m & $\sim$18.2 & \citet{schuller_2009} \\
{\it Herschel} Infrared Galactic Plane Survey (Hi-GAL) & 160, 250, 350, and 500 $\mu$m & $\sim$12, 18, 25, and 37 &\citet{Molinari10a}\\
Galactic Legacy Infrared Mid-Plane Survey Extraordinaire (GLIMPSE) & 3.6, 4.5, 5.8 $\mu$m & $\sim$2 & \citet{Benjamin}\\
Vista Variables in the V{\'i}a L{\'i}actea (VVV) & 1.6 and 2.2 $\mu$m & $\sim$0.8 & \citet{Minniti_2010}\\
Two Micron All Sky Survey (2MASS) & 1.6 and 2.2 $\mu$m & $\sim$2.5 & \citet{Skrutskie_2006}\\
\hline \hline
\end{tabular}
\label{tab_data_g321}
\end{table*}
%

\begin{table*}
\centering
\caption{Physical parameters of the {\htwo} regions.}
\begin{tabular}{ccccc}
\hline \hline
 {\htwo} Regions &   $R_{\rm HII}$ (pc) &  $S_\nu$ (mJy) &  log($N_{\rm uv}$ [s$^{-1}$])& $t_{\rm {dyn}}$ (Myr)\\
\hline 
 {\htwo}-1 &  0.05 &          0.34 &    44.02   &0.02 \\
 {\htwo}-2 &  0.22 &         13.32 &    45.61   &0.12 \\
 {\htwo}-3 &  0.34 &         25.67 &    45.89   &0.22 \\
 {\htwo}-4 &  0.09 &          6.17 &    45.27   &0.03 \\
\hline \hline
\label{tab1_htwo}
\end{tabular}
\end{table*}

\section{Results}
\label{sec_results_G321}
\subsection{Physical environment of G321}
\label{sec_phy_env_G321}
Star-forming activity is closely related to the density and temperature of the medium. Figures~\ref{G321_herschel_maps}b and~\ref{G321_herschel_maps}c present the $N$(H$_2$)) and $T_\mathrm{d}$ maps for our target area, respectively, with a resolution of about 12\as. These maps were derived from {\it Herschel} images using the tool {\it hires} \citep{getsf_2022}.
The $N$(H$_2$) map clearly shows the presence of filaments in the area of our interest. The $N$(H$_2$) value toward the position of the ATLASGAL clumps is above 2 $\times$ 10$^{22}$ cm$^{-2}$, while the filaments are of relatively lower column density. The dust temperature toward the filamentary structures observed in the $N$(H$_2$) map is about 15 K. A relatively higher $T_\mathrm{d}$ (i.e., $\sim$20 K) is observed toward the high-column density regions, which are shown with yellow arrows in Figure~\ref{G321_herschel_maps}b. The SUMSS radio continuum emission is detected exclusively toward these regions. On the other hand, the high-column density regions, which are marked with cyan arrows in Figure~\ref{G321_herschel_maps}b, are relatively cooler, and their $T_\mathrm{d}$ is about 10 K. The availability of the MeerKAT 1.28 GHz high-resolution radio continuum image reveals the presence of four compact {\htwo} regions (see {\htwo}-1--{\htwo}-4 in Figure~\ref{G321_herschel_maps}c). Their effective radii ($R_{\rm HII}$), total flux density ($S_\nu$), and dynamical timescale ($t_{\rm {dyn}}$) are listed in Table~\ref{tab1_htwo} (see Appendix~\ref{sec_appendix_sptypeHII} for details).  

The large-scale molecular gas distribution toward G321 is presented in Figure~\ref{321_molecular_maps} using SEDIGISM $^{13}$CO($J$ = 2--1) data. Figure~\ref{321_molecular_maps}a is the $^{13}$CO($J$ = 2--1) integrated emission (i.e., the moment-0) map for the velocity ($v_{\mathrm {lsr}}$) range [$-$38.25, $-$27.25] {\kps}, showing the morphology of the molecular cloud in the plane of sky. The dotted rectangle highlights the area covered in Figure~\ref{G321_herschel_maps}a for {\it Herschel} continuum images. 
The peak emission map is displayed in Figure~\ref{321_molecular_maps}b. Both Figures~\ref{321_molecular_maps}a and \ref{321_molecular_maps}b reveal intense molecular emission toward the position of the ATLASGAL clumps and the presence of several filamentary structures similar to those observed in the dust continuum images in Figure~\ref{G321_herschel_maps}a.
The intensity-weighted velocity (i.e., the moment-1) map and the velocity map corresponding to the peak intensity (referred to as the `peak velocity map') are presented in Figures~\ref{321_molecular_maps}c and \ref{321_molecular_maps}d, respectively. The moment-1 map displays the spatial distribution of overall gas velocity and provides the first hint about the presence of two velocity components, which appear in blue and red. The peak velocity map indicates that the filamentary structures observed in the peak intensity map have quite similar velocities across their length. This feature has been well reflected in Figure~\ref{321_channel_maps}, which presents the moment-0 maps of {\tco} emission at $v_{\mathrm {lsr}}$ from $-$38.25 to $-$27.25 {\kps}, with an interval of about 0.75 {\kps}. This image reveals how the gas velocity changes over cloud structures. The narrow integration range (i.e., $<$ 1 {\kps}) ensures that the gaseous structures observed in Figure~\ref{321_channel_maps} are velocity-coherent features of the molecular cloud.
\subsection{Detection of filament skeletons and the distribution of color excess sources}
\label{sec_fil_G321}
The visual inspection of the {\it Herschel} continuum images in Figure~\ref{G321_herschel_maps}a and the molecular gas emission in Figures~\ref{321_molecular_maps} and \ref{321_channel_maps} indicate the presence of filaments in our target area. To identify the filament skeletons, we utilized {\it getsf} \citep{getsf_2022} on the $^{13}$CO($J$ = 2--1) moment-0 map, shown in Figure~\ref{321_molecular_maps}a. The {\it getsf} tool requires four inputs to function: the distance to the source, the angular resolution of the image, and a rough estimation of the widths of the largest filament and core. In this case, the distance of the source and the angular resolution of the image are 1.98 kpc \citep{Atlasgal} and 36{\as}, respectively. Through visual inspection of the moment-0 map in DS9{\footnote{https://sites.google.com/cfa.harvard.edu/saoimageds9}}, we estimated the widths of the largest filament and core to be 260{\as} and 160{\as}, respectively. The last two inputs are required to set a cutoff during successive unsharp masking to prevent decomposition to a very large scale \citep[][]{getsf_2022}. The {\it getsf}-identified filament skeletons (on the scale of 204{\as}) are highlighted in Figure~\ref{321_filament_maps}a over the $^{13}$CO($J$ = 2--1) moment-0 map. Several parsec-scale filaments are found to converge at two distinct junctions, characterized by intense molecular emission and high $N$(H$_2$) values (see the arrows in Figure~\ref{321_filament_maps}a), which perfectly match the definition of HFS \citep[see][]{myers_2009}. Therefore, at least two HFSs, named HFS-1 and HFS-2, are detected using {\it getsf} in our target area. The hubs corresponding to HFS-1 and HFS-2 are labeled as Hub-1 and Hub-2, respectively, in Figure~\ref{321_filament_maps}a. The detection of two HFSs toward G321 is an interesting finding considering the fact that there are only a few sources reported in the literature with more than one large-scale HFS ($>$ 1 pc) in the same star-forming site, such as IC 5146 \citep{Wang_2019,Dewangan2023ApJ}, G45.3+0.1 \citep{Bhadari_2022ApJ}, and Galactic `Snake' \citep{Dewangan_2024MNRAS}.

The high-density regions are susceptible to star-forming activity, which can be inferred from the detection of embedded protostars or Class\,{\sc I} young stellar objects (YSOs), which show infrared excess due to their envelopes and the dusty circumstellar disks \citep{Povich_2011ApJS,Sharma_2017MNRAS}. We have utilized {\it Spitzer} photometric data (at 3.6--5.8 {\micro}) to detect {\cone} in our target site, considering point-like sources with a photometric magnitude error of 0.2 mag or less. Previous studies of \citet{hartmann_2005} and \citet{getman_2007} showed that {\cone} can satisfy the color conditions: [4.5] $-$ [5.8] $\geq$ 0.7 and [3.6] $-$ [4.5] $\geq$ 0.7. 
The positions of 61 Class\,{\sc I} YSO candidates are displayed on the $^{13}$CO moment-0 map in Figure~\ref{321_filament_maps}b, within the yellow-dotted region. Several YSO candidates are detected toward both hubs (Hub-1 and Hub-2; see Figure~\ref{321_filament_maps}a) and the filaments. Interestingly, we observe a relatively large number of YSO candidates toward Hub-1 compared to Hub-2. 
In Figure~\ref{321_filament_maps}c, we have visually marked the extent of Hub-1 and Hub-2 with two blue circles of sky-projected radii about 2.1 and 1.4 pc, respectively. Hub-1 and Hub-2 are zoomed-in using the VVV $K_{\mathrm{s}}$ band image in the top-left and bottom-right insets, respectively. The superior spatial resolution and sensitivity of the VVV $K_{\mathrm{s}}$ band image reveal many point-like sources compared to the {\it Spitzer} images (not included here) for both hubs. The red rectangles overplotted on the VVV $K_{\mathrm{s}}$ band images in the insets of Figure~\ref{321_filament_maps}c highlight the sources with color excess, where $H - K_{\mathrm{s}} > 1.8$. This cutoff value has been obtained from the color-magnitude analysis of a nearby control field (size {\simi}4{\rlap{$'$}}.5 $\times$ 4{\rlap{$'$}}.5, centred at $l = $ 321{\rlap{$^\circ$}} .724 and $b = -$0{\rlap{$^\circ$}} .225). This cutoff value of 1.8 was utilized to detect the color excess sources for the MSFR toward $l =$ 345{\rlap{$^\circ$}} .5 and $b =$ 0{\rlap{$^\circ$}} .3 in the study of \citet{dewangan_2018_dec}. The better photometric depth of the VVV data allows the detection of 140 color excess sources toward Hub-1 and 32 for Hub-2. Since all these NIR color-excess sources are distributed toward regions of intense molecular emission, it is unlikely that they are foreground or background sources; therefore, they can be considered YSO candidates. 
The use of 2MASS data for bright sources adds 6 more unique color excess sources for Hub-1, while it does not contribute any additional color excess sources for Hub-2. 

\subsection{The physical properties of the molecular cloud, hubs, and the filaments}
In this section, we estimate the mass of the molecular cloud, hubs, and the filaments. We also determine the line-mass of the filaments and the mass accretion rate along them toward their hubs in HFS-1 and HFS-2.
 
\subsubsection{Mass estimation using SEDIGISM $^{13}$CO($J$ = 2--1) and C$^{18}$O($J$ = 2--1) data}
\label{sec_mass_G321}

The $^{13}$CO column density (i.e., $N$($^{13}$CO)) can be calculated using the equation form \citet{Mangum_2015_pasp},
\begin{eqnarray}
N(^{13}\mathrm{CO})&=&\frac{3h}{8\pi^3\mu_{\rm {dm}}^2S}\frac{Q_\mathrm{rot}}{g_{J}}
\frac{\mathrm{exp}\left(\frac{E_\mathrm{up}}{k T_\mathrm{ex}}\right)}{\mathrm{exp}\left(\frac{h\nu}{k T_\mathrm{ex}}\right) - 1} \label{calculate_N13CO}\\
&&\times\frac{1}{J(T_\mathrm{ex})-J(T_\mathrm{bg})}\frac{\tau_{13}}{1-\mathrm{exp}(-\tau_{13})}\int T_\mathrm{mb}~dv \nonumber,
\end{eqnarray}
with an assumption that the system is in local thermodynamic equilibrium. Here, $T_\mathrm{ex}$ and $T_{\rm bg}$ stand for the excitation temperature related to our target source and the cosmic microwave background temperature, respectively. $T_\mathrm{mb}$ represents the main beam temperature of the line emission.   
$J(T)$ is the Rayleigh-Jeans equivalent temperature, which is defined as $({h\nu/k})/({\mathrm{exp}(h\nu/k T)-1})$. The energy of the upper state, its degeneracy, and the line strength are denoted with $E_\mathrm{}$, $g_\mathrm{J}$, and $S$, respectively. For the $J = 2-1$ transition, $g_\mathrm{J} = 2J_u+1 = 5$ and $S = {J_u}/(2J_u+1) = {2}/{5}$. The rotational partition function ($Q_\mathrm{rot}$) is approximated to be ${KT}/{hB} + {1}/{3}$ \citep[][]{McDowell_1988,Yuan_2016_apj}, where $B$ is the the rotational angular momentum constant. From the Jet Propulsion Laboratory (JPL) spectroscopic database{\footnote{https://spec.jpl.nasa.gov/ftp/pub/catalog/catdir.html}} \citep{Pickett_1998}, we obtained $B$ = 55101.011 MHz and the dipole moment, $\mu_{\rm {dm}}$ = 0.11046 Debye for $^{13}$CO molecule.
The optical depth for the {\tco} emission ($\tau_{13}$) can be estimated
from the ratio of the line peak intensities of {\tco} and {\cetno} data \citep[see][]{Miettinen_2012AA,Mangum_2015_pasp,Liu_2020_apj},
 \begin{equation}
    \frac{T_{\rm mb}{\rm (^{13}CO)}}{T_{\rm mb}{\rm (C^{18}O)}}\approx\frac{1-{\rm exp} (-\tau_{13})}{1-{\rm exp} (-\tau_{13}/R)}, 
    \label{eq-tau}
 \end{equation}
where, $R$ \citep[= 7.4;][]{Areal_2018} is the isotopic ratio of $^{13}$CO and C$^{18}$O.
The $\tau_{13}$ values are numerically calculated using Equation~\ref{eq-tau} for pixels with peak intensities greater than 3$\sigma$, where 1$\sigma$ = 0.35 K for both the $^{13}$CO and C$^{18}$O data (see Section~\ref{sec_obser_G322}). The $\tau_{13}$ map and its statistics are included in Appendix~\ref{sec_appendix_tau13} (see Figure~\ref{tau13andNH2}a). Then, we obtained the $N(^{13}\mathrm{CO})$ values using $T_{\rm bg} = 2.73$ K, $T_\mathrm{ex} =$ 15 (20) K, the integrated {\tco} emission, and the $\tau_{13}$ values in Equation~\ref{calculate_N13CO}. The $N(^{13}\mathrm{CO})$ values are then converted to $N(\mathrm{H_2})$ for ${N(\mathrm{H_2})}/{N(\mathrm{^{13}CO})}$ = $7 \times 10^5$ \citep[e.g.,][]{Frerking_1982apj}. The $N(\mathrm{H_2})$ map and its statistics for $T_\mathrm{ex} =$ 15 K can be found in Figure~\ref{tau13andNH2}b in Appendix~\ref{sec_appendix_tau13}. 
Figure~\ref{tau13andNH2} shows that estimations of $\tau_{13}$ and the corresponding $N(\mathrm{H_2})$ are unavailable for many pixels due to the absence of {\cetno} emission above 3$\sigma$. Hence, we have established the relationship, $N(\mathrm{H_2}) \sim1.173(1.176) \times 10^{21}$ cm$^{-2}$ (K {\kps})$^{-1}$ for $T_\mathrm{ex} = 15(20)$ K, by comparing the total $N(\mathrm{H_2})$ and {\tco} moment-0 values for pixels where $N(\mathrm{H_2})$ values are available. This relationship between $N(\mathrm{H_2})$ and {\tco} moment-0 value is in good agreement with the result of \citet{schuller17} and is used to derive the $N(\mathrm{H_2})$ values for the entire molecular cloud. However, it is important to note that our estimated $N(\mathrm{H_2})$ values are uncertain by a factor of a few, which will propagate to all the physical parameters calculated based on this $N(\mathrm{H_2})$ values.

The knowledge of $N(\mathrm{H_2})$, the mean molecular weight ($\mu$; assumed to be 2.8\thinspace $m_\mathrm{H}$ from \citet{Kauffmann_2008}; where $m_\mathrm{H}$ is the mass of a hydrogen atom), and the distance of the the molecular cloud (i.e., $d = 1.98$ kpc) allows for the calculation of its total mass. The computed mass will have an uncertainty of a factor of a few, mainly due to the uncertainty in $N(\mathrm{H_2})$.
For the moment-0 values above 3$\sigma$ (see Figure~\ref{321_molecular_maps}), the total mass of the molecular cloud is about $7.93 (7.91) \times 10^4$ {\Msolar} for $T_\mathrm{ex} =$ 15 (20) K, respectively. Similarly, the total mass of Hub-1 and Hub-2 are estimated to be nearly $1.60 \times 10^4$ {\Msolar} and $3.81 \times 10^3$ {\Msolar}, respectively. There is no significant difference in the masses of the hubs for $T_\mathrm{ex} =$ 15 and 20 K, up to two decimal places.
The red circles{\footnote{By comparing the filament F6 observed in Figures~\ref{321_channel_maps}k and \ref{321_channel_maps}l, we noticed that the {\it getsf}-identified skeleton for F6 is not accurate. Consequently, we manually selected the first four circular regions of F6 toward Hub-2.}} with a diameter of about 12 pixels (i.e., {\simi}1.1 pc; see Figure~\ref{321_filament_maps}c) along the filaments are utilized to extract the velocity profile along the length of the filaments, which are detailed in Section~\ref{sec_mass_acc_G321}. 
We calculated the mass of these filaments (i.e., F1--F7) for $T_\mathrm{ex} = 15$ K, using the total {\tco} moment-0 values, covered with the red circles. The length ($L^{\mathrm{Fil}}$), mass ($M^{\mathrm{Fil}}$), and line-mass (i.e., $(M/L)^{\mathrm{Fil}}$) of the filaments are listed in Table~\ref{tab_velograd}. Based on the $(M/L)^{\mathrm{Fil}}$ values, the stability of the filaments is discussed in Section~\ref{sec_role_fil_G321}.


\begin{table*}
	\centering
	\caption{Physical properties of the filaments. 
	}
	\begin{tabular}{ccccccccc} 
		\hline \hline
		IDs & Associated with & $L^{\rm Fil}$ &$M^{\rm Fil}$ & $(M/L)^{\rm Fil}$ &$\sigma^{\rm Fil}_{\rm avg}$&$\nabla$V$_{||}^{\rm obs}$ & $\dot{M}_{||}$ &  $\frac{(M/L)^{\rm Fil}}{(M/L)_{\rm crit}^{\rm {tur}}}$  \\
		    &   & (pc) &($\times$ 10$^3$$M_{\odot}$)&($\times$ 10$^2$$M_{\odot}$ pc$^{-1}$) & ({\kps}) &({\kps}pc$^{-1}$)      & ($\times 10^{-3}$$M_{\odot}$ yr$^{-1}$) & \\

		\hline
		F1 & Hub-1      & 12.08 & 6.88 & 5.70& 1.11 & 0.12 &   8.43  & 1.0\\
        F2 & Hub-1      &  3.20 & 2.84 & 8.88& 1.24 & 0.23 &   6.67  & 1.2\\
        F3 & Hub-1      &  4.50 & 3.55 & 7.89& 1.25 & 0.31 &  11.25  & 1.1\\
        F4 & Hub-1 \& 2 &  4.35 & 4.53 &10.41& 1.29 & --- &   ---    & 1.3\\
        F5 & Hub-2      &  2.23 & 1.58 & 7.09& 1.05 & 0.10 &   1.61  & 1.4\\
        F6 & Hub-2      &  8.92 & 2.78 & 3.12& 1.27 & 0.09 &   2.56  & 0.4\\
        F7 & Hub-2      &  4.90 & 2.90 & 5.92& 1.79 & --- &   ---    & 0.4\\
	\hline \hline
	\end{tabular}
 \label{tab_velograd}
\end{table*}

\subsubsection{Estimation of the mass accretion rate through the filaments to the hubs}
\label{sec_mass_acc_G321}
The mass accretion rate ($\dot{M}_{||}$) along the filament is a key parameter for the HFSs. Hence, we calculated $\dot{M}_{||}$ using the equation from \citet{kirk_2013},
\begin{equation}
   \dot{M}_{||}=\frac{\nabla V_{||}^{\mathrm{obs}} M^{\mathrm{Fil}}}{\mathrm{tan}(\alpha)}
    \label{eq-acc-rate},
\end{equation}
where, $\nabla V_{||}^{\mathrm{obs}}$ represents the observed velocity gradient along the filament, and $\alpha$ is its angle relative to the plane of the sky. 
The average velocity for the circular regions of filaments F1–F7 (see Figure~\ref{321_filament_maps}c) is estimated by fitting Gaussians to their velocity profiles and is shown as a function of distance from their hubs in Figure~\ref{321_velocity_gradient}. We then applied a linear fit (see the red lines) to the average velocity distribution along the length of the filaments to estimate $\nabla V_{||}^{\mathrm{obs}}$. The velocity dispersion for these regions is indicated using a color scale, and the average velocity dispersion ($\sigma^{\rm Fil}_{\rm avg}$) for F1--F7 is listed in Table~\ref{tab_velograd}.
The physical association of filament F4 with both Hub-1 and Hub-2, along with the presence of a dense fragment (as detailed in Section~\ref{sec_ATLASGAL_ALMA_G321}), complicates the velocity distribution along the filament, making it impractical to apply the simple mass accretion model proposed by \citet{kirk_2013}. The moment-0 maps for narrow velocity integration ranges (see Figure~\ref{321_channel_maps}) do not show a clear signature of filament F7; hence, the detection of this filament is highly doubtful. Therefore, the velocity gradients for filaments F4 and F7 have not been calculated. 
The finite $L^{\mathrm{Fil}}$ and $\nabla V_{||}^{\mathrm{obs}}$ indicate that $\alpha$ is neither close to $0^\circ$ nor $90^\circ$ for the filaments F1, F2, F3, F5, and F6. Hence, assuming $\alpha = 45^\circ$ and using $M^{\mathrm{Fil}}$ and $\nabla V_{||}^{\mathrm{obs}}$ in Equation~\ref{eq-acc-rate}, we obtained their $\dot{M}_{||}$ values, which are listed in Table~\ref{tab_velograd}. The $\dot{M}_{||}$  varies from $1.61\times10^{-3}$ to $11.25\times10^{-3}$ $M_{\odot}$ yr$^{-1}$ for the filaments F5 and F3, respectively. The combined $\dot{M}_{||}$ in Hub-1 through filaments, F1, F2, and F3 is about $26.35\times10^{-3}$ $M_{\odot}$ yr$^{-1}$. However, for Hub-2, combined $\dot{M}_{||}$ through filaments, F5, and F6 is about $4.17\times10^{-3}$ $M_{\odot}$ yr$^{-1}$. Our estimated mass accretion rate can vary by a factor of 0.58 to 1.73 for $\alpha = 30^\circ$ to $60^\circ$. It is important to note that the {\it getsf}-estimated average width of the filaments is about 1.8 pc. However, to avoid contamination from the surrounding molecular gas and to better determine the average velocity, we selected circular regions with a relatively smaller diameter ({\simi}1.1 pc). This same width was used to calculate the total mass of the filaments; hence, our reported values of $M^{\mathrm{Fil}}$ and $\dot{M}_{||}$ represent lower limits.
It is important to note that the analytical model by \citet{kirk_2013} is simplistic, assuming filaments as cylinders of uniform density and neglecting the effects of core formation and stellar feedback. Hence, the varying density of the filament, the formation of its core, and the influence of stellar feedback can impact the velocity profile along the length of the filament. Furthermore, any deviation from the ideal cylindrical shape will result in different projection angles at various points along the filament, thus affecting its velocity profile. These factors prevent a perfect linear fit of the average velocity plot, as seen in Figure~\ref{321_velocity_gradient}. Therefore, a better model is necessary for more precise results.

\subsection{Study of the hierarchical structures form clump- to the core-scale}
\label{sec_ATLASGAL_ALMA_G321}
Star-forming molecular clouds often exhibit hierarchical density structures. To reveal the same in our target site, we performed a dendrogram analysis using the Python-based tool {\it astrodendro}\footnote{https://dendrograms.readthedocs.io/en/stable/index.html}, which can be applied to astronomical 2D images as well as 3D (position-position-velocity) data cubes \citep[][]{Rosolowsky_2008ApJ,Goodman_2009Natur,Burkhart_2013ApJ}. 
For 2D images, {\it astrodendro} requires three input parameters: 1. {\it min\_value}, the minimum flux density for pixels to be included in a structure, 2. {\it min\_delta}, the minimum peak density difference required between two potential structures to consider them separate entities, and 3. {\it npix}, the minimum number of pixels necessary to form the smallest possible structure. The physical interpretation of these input parameters is illustrated in detail using a cartoon diagram in Figure 6 of \citet{Chen_2018ApJ}. Based on the inputs, {\it astrodendro} provides a hierarchical tree structure composed of branches and leaves. Branches, which are larger and fainter structures, are positioned lower in the tree and can break down into new branches and leaves. Leaves are small, bright structures at the tips of the tree that do not subdivide.
\subsubsection{Clump-scale hierarchical structures in ATLASGAL 870 {\micron} continuum image}
\label{dendrogram_ATLASGAL}
We performed the dendrogram analysis on the ATLASGAL 870 $\mu$m continuum image to study clump-scale ({\simi}1 pc) structures in the rectangular area highlighted in Figure~\ref{321_channel_maps}g. The input parameters for {\it astrodendro} were {\it min\_value} = 3$\sigma$ and {\it min\_delta} = 1$\sigma$ ({\simi}60 mJy beam$^{-1}$). The {\it npix} value was set to be 22, corresponding to twice the beam area of the ATLASGAL data. The hierarchical structures, i.e., the branch and leaves identified in the ATLASGAL continuum image, are highlighted using red and cyan contours in Figure~\ref{321_ATLASGAL_154}a, respectively. The radii of these structures are calculated as $R = \sqrt{A/\pi}$, where $A$ is the exact area. After correcting $R$ for the ATLASGAL beam size, we have listed the deconvolved radii as $R_{\mathrm{eff}}$ in Table~\ref{tab_atlasgal_tree}, along with the values of $A$ and $S_\nu$. $R_{\mathrm{eff}}$ ranges from 0.24 to 0.92 pc. The extent of the hubs (as shown in Figure~\ref{321_filament_maps}c) is marked using dotted circles in Figure~\ref{321_ATLASGAL_154}a. Hub-1 hosts the largest structure (i.e., ATLASGAL Branch (ATB)-1; $R_{\mathrm{eff}} = 0.92$ pc), which is further fragmented into two leaves, namely ATLASGAL Leaf (ATL)-3 and ATL-4. ATL-6 resides within Hub-2, and ATL-5 is projected over filament F4, which connects both the hubs. We have estimated the average $T_\mathrm{d}$ for the branch and the leaves (for their area depicted in Figure~\ref{321_ATLASGAL_154}a) from the {\it Herschel} dust temperature map, which are specified in Table~\ref{tab_atlasgal_tree}.

The knowledge of $S_\nu$, $T_{\mathrm d}$, and $d$ of these structures allows the estimation of their total mass using the formula from \citet{Hildebrand_1983}, 
\begin{equation}
 M= \frac{S_\nu~d^2~R_t}{B_{(\nu,T_{\mathrm d})}~k_\nu},
\label{eq-mass}
\end{equation}
where, $B_{(\nu,T_{\mathrm d})}$, $R_t$, and  $k_\nu$ are Planck function, gas-to-dust mass ratio, and dust absorption coefficient, respectively. 
For ATLASGAL 870 $\mu$m (i.e., 344.59 GHz) emission, $k^{870 \mu m}_{\rm 344.59~GHz}$ is about 1.54 cm$^{2}$ g$^{-1}$, derived from the relation $k_\nu = 10\left({\nu}/{1.2~[{\rm THz}]}\right)^{1.5}$ \citep{Hildebrand_1983,Li_2020ApJ}. Then, using average $T_{\mathrm d}$ values, $d=1.98$ kpc, and $R_t=100$ \citep[e.g.,][]{Weingartner_2001ApJ,Mutie_2021} in Equation~\ref{eq-mass}, we estimated the mass of the dendrogram structures. The $M$--$R_{\mathrm{eff}}$ plot for the {\it astrodendro}-identified structural components are shown in Figure~\ref{321_ATLASGAL_154}b. The blue dashed line indicates the Kauffmann \& Pillai (hereafter, KP-10) condition for MSF, i.e., $M(R) > 870$ {\Msolar} $({R}/{\mathrm{pc}})^{1.33}$ \citep[][]{Kauffmann_2010ApJ}. For the dust opacity value used in this paper, the mass coefficient provided by \citet[][]{Kauffmann_2010ApJ} is reduced by a factor of 1.249. Hence, the modified KP-10 (hereafter, mKP-10) condition for MSF is $M(R) > 697$ {\Msolar} $({R}/{\mathrm{pc}})^{1.33}$ \citep[][]{Kauffmann_2010ApJ}. The white region above the gray-shaded area in Figure~\ref{321_ATLASGAL_154}b satisfies the mKP-10 condition.
Branch-1 and its associated ATL-3 significantly exceed the mKP-10 limit, while ATL-4, ATL-5, and ATL-6 are positioned near the threshold for MSF. On the other hand, ATL-2 significantly falls below the mKP-10 limit. 
Furthermore, assuming the structures to be spherical, we can derive their density and column density using the formulas: $n = {3M}/{4\pi \mu \mathrm{m_H} R_{\mathrm{eff}}^3}$ \citep{Li_2020ApJ} and $N(\mathrm{H_2}) = {M}/{\pi \mu \mathrm{m_H} R_{\mathrm{eff}}^2}$ \citep{Bhadari_2023MNRAS}, respectively.
The mass, density, and column density of all the dendrogram structures are listed in Table~\ref{tab_atlasgal_tree}. These structures have a density of about 10$^4$ cm$^{-3}$, with $N(\mathrm{H_2}) > 10^{22}$ cm$^{-2}$. The typical error in the mass estimation is about 20\%, considering the uncertainties in $T_{\mathrm d}$, $S_\nu$, and $d$ are about 10\%, 15\% \citep{schuller_2009}, and 6\% \citep{Atlasgal}, respectively. The density and column density have similar uncertainties as the mass of the structures. Given that $k_{\nu}$ and $R_t=100$ can have significant uncertainties \cite[see][]{Li_2020ApJ}, the 20\% error should be considered a lower limit, and the actual error could be several times larger. 

\subsubsection{Core-scale hierarchical structures in ALMA Band-7 data}
\label{dendrogram_ALMA} 
The area covered by the blue circle within ATL-3 (see Figure~\ref{321_ATLASGAL_154}a) is zoomed-in in Figure~\ref{321_ALMA_images}a using ALMA Band-7 continuum data (beam size {\simi}4{\rlap {\as}} .9 $\times$ 3{\rlap {\as}} .1). A single structure is traced in the ALMA data at this angular resolution above the 3$\sigma$ limit (where 1$\sigma$ {\simi}40 mJy beam$^{-1}$), which is highlighted with the yellow contour. 
The star-forming activity of this structure is revealed through the detection of Class I YSO candidates, as well as 22 GHz H$_2$O and Class I 95 GHz CH$_3$OH maser emissions toward it (see Figure~\ref{321_ALMA_images}a). 
The total flux density for the structure is calculated to be about 4.43 Jy. Following the formulas described in Section~\ref{dendrogram_ATLASGAL}, we calculated the basic physical parameters for this structure, which are $R_{\mathrm{eff}}$ \simi$0.07$ pc, $M$ \simi$139$ \Msolar, $n > 10^{6}$ cm$^{-3}$, and $N(\mathrm{H_2}) > 10^{23}$ cm$^{-2}$. Here, $R_{\mathrm{eff}}$ is the geometric mean of the semi-major and semi-minor axis corresponding to the exact area of the structure, after being corrected for the beam size. For the mass estimation, we used $k^{880 \mu m}_{\rm 341.06~GHz}$ = 1.52 cm$^{2}$ g$^{-1}$ and the average dust temperature of ATL-3 (i.e., $T_{\rm{d}}$ = 18 K).
The availability of ALMA Band-7 continuum image with a better resolution (beam size $\sim$0{\rlap {\as}} .33 $\times$ 0{\rlap {\as}} .16) allowed us to further zoom-in on the area covered by the red dotted circle in Figure~\ref{321_ALMA_images}a. The zoomed-in image is shown in Figure~\ref{321_ALMA_images}b. To explore the hierarchical structures at the core scale (i.e., $\lesssim$ 0.01 pc), we utilized {\it astrodendro} on the ALMA Band-7 high-resolution image. The input parameters for the {\it astrodendro} analysis were {\it min\_value} = 3$\sigma$, {\it min\_delta} = 1$\sigma$ ({\simi}2 mJy beam$^{-1}$), and {\it npix} = 138, corresponding to twice the beam area of the ALMA data.
Similar to Figure~\ref{321_ATLASGAL_154}a, the {\it astrodendro}-identified branches and leaves are highlighted in the ALMA Band-7 continuum image using red and cyan contours, respectively (see Figure~\ref{321_ALMA_images}b). 
The positions of the Class I YSO candidates, 22 GHz H$_2$O, and 95 GHz Class I CH$_3$OH maser emissions are overlaid in Figure~\ref{321_ALMA_images}b. This overlay emphasizes the spatial association between the ALMA branches and leaves, the YSO candidates, and the maser emissions.
The dendrogram tree obtained from the {\it astrodendro} analysis of the ALMA data is shown in Figure~\ref{321_ALMA_images}c. This tree illustrates how the branches are further fragmented into branches and/or leaves. We have estimated physical parameters for the branches and leaves using the same input parameters as earlier for the low-resolution ALMA data, and these parameters are listed in Table~\ref{tab_alma_tree}. The mass of the branches varies from 8 to 33 \Msolar, while the mass of the leaves is less than 10 \Msolar. The density of these structures is about 10$^7$ cm$^{-3}$, and $N(\mathrm{H_2}) > 10^{24}$ cm$^{-2}$. The uncertainties in $T_{\mathrm d}$, $S_\nu$, $d$, $k_{\nu}$, and $R_t$ propagate to a minimum uncertainty of {\simi}50\% in the estimation of mass, density, and column density of the ALMA cores  \citep[see][for details]{Sanhueza_2017ApJ,Barnes_2021MNRAS}. The $M$--$R_{\mathrm{eff}}$ plot of the {\it astrodendro}-identified structural components from the ALMA low- and high-resolution data is shown in Figure~\ref{321_ALMA_images}d, along with their host ATLASGAL Branch (i.e., ATB-1) and Leaf (i.e., ATL-3). All the structures, from ATB-1 to the ALMA leaves/cores, satisfy the mKP-10 condition for MSF.

\begin{table*}
\caption{Physical parameters for the {\it astrodendro}-identified hierarchical structures in ATLASGAL 870 {\micro} continuum image.}
\begin{tabular}{ccccccccc}
\hline \hline
IDs &   Type & Exact Area                     & $R_{\mathrm{eff}}$ & $S_\nu$ &  $T_\mathrm{d}$ & Mass       &      Density                    &  log($N$(H$_2$) [cm$^{-2}$])\\
    &        & ($\times$ 10$^{3}$ arcsec$^2$) &  (pc)              &   (Jy)  & (K)             &({\Msolar}) &     ($\times$ 10$^3$ cm$^{-3}$) &  \\
\hline
   1 & Branch &        29.1 &  0.92 &     46.10 &    17 & 1532 &      6.8 &   22.41 \\
   2 &   Leaf &         2.2 &  0.24 &      2.09 &    17 &   70 &     17.3 &   22.23 \\
   3 &   Leaf &         7.0 &  0.44 &     24.64 &    18 &  749 &     29.6 &   22.73 \\
   4 &   Leaf &         2.3 &  0.25 &      4.15 &    15 &  168 &     38.7 &   22.59 \\
   5 &   Leaf &         4.7 &  0.36 &      5.52 &    14 &  251 &     18.6 &   22.44 \\
   6 &   Leaf &         8.0 &  0.48 &      7.27 &    14 &  332 &     10.7 &   22.32 \\

\hline \hline
\end{tabular}
\label{tab_atlasgal_tree}
\end{table*}

\begin{table*}
\centering
\caption{Physical parameters for the {\it astrodendro}-identified hierarchical structures in ALMA Band-7 continuum image. Dust temperature for ATL-3 (i.e., $T_{\rm d}$ = 18 K) is utilized for the determination of mass of all the structures (i.e., branches and leaves).}
\begin{tabular}{cccccccc}
\hline \hline
IDs &   Type & Exact Area &  $R_{\mathrm{eff}}$  & $S_\nu$& Mass &      Density &  log($N$(H$_2$) [cm$^{-2}$])\\
    &   &(arcsec$^2$) &  ($\times$ 10$^{-3}$ pc)  & (Jy) &({\Msolar}) &     ($\times$ 10$^6$ cm$^{-3}$) &  \\
\hline
   1 &   Leaf &        0.67 &   4.3 &      0.10 &   3.0 &     14.0 &   24.39 \\
   2 &   Leaf &        0.38 &   3.2 &      0.05 &   2.0 &     16.9 &   24.34 \\
   3 &   Leaf &        0.17 &   1.9 &      0.03 &   1.0 &     45.8 &   24.57 \\
   4 & Branch &        5.68 &  12.9 &      1.07 &  33.0 &      5.4 &   24.46 \\
   5 &   Leaf &        1.10 &   5.6 &      0.18 &   6.0 &     11.1 &   24.40 \\
   6 &   Leaf &        0.24 &   2.4 &      0.05 &   2.0 &     42.3 &   24.61 \\
   7 & Branch &        1.78 &   7.1 &      0.50 &  16.0 &     15.0 &   24.64 \\
   8 &   Leaf &        0.41 &   3.3 &      0.15 &   5.0 &     46.4 &   24.79 \\
   9 &   Leaf &        0.75 &   4.5 &      0.11 &   3.0 &     12.3 &   24.36 \\
  10 & Branch &        0.94 &   5.1 &      0.26 &   8.0 &     21.3 &   24.65 \\
  11 &   Leaf &        0.27 &   2.6 &      0.07 &   2.0 &     44.5 &   24.67 \\
  12 &   Leaf &        0.46 &   3.5 &      0.15 &   5.0 &     36.5 &   24.72 \\
  13 & Branch &        3.71 &  10.4 &      0.69 &  22.0 &      6.7 &   24.46 \\
  14 &   Leaf &        1.43 &   6.4 &      0.29 &   9.0 &     12.2 &   24.50 \\
  15 & Branch &        2.39 &   8.3 &      0.52 &  16.0 &     10.0 &   24.53 \\
  16 &   Leaf &        0.92 &   5.1 &      0.22 &   7.0 &     18.7 &   24.59 \\
  17 &   Leaf &        0.31 &   2.8 &      0.07 &   2.0 &     31.6 &   24.56 \\
  18 &   Leaf &        0.65 &   4.2 &      0.10 &   3.0 &     15.0 &   24.41 \\
  19 &   Leaf &        0.43 &   3.4 &      0.10 &   3.0 &     26.5 &   24.57 \\

\hline \hline
\end{tabular}
\label{tab_alma_tree}
\end{table*}
\subsection{Identification of different velocity components and their spatial distribution}
\label{sec_mol_G321}
As mentioned earlier in Section~\ref{sec_phy_env_G321}, the moment-1 map of {\tco} emission indicates the presence of two distinct velocity components, represented in colors blue and red (see Figure~\ref{321_molecular_maps}c). To identify the different velocity components of the molecular cloud, the Galactic longitude--velocity (i.e., {\it l}--{\it v}) diagram is produced using the $^{13}$CO($J$ = 2--1) data, which is displayed in Figure~\ref{321_twocolor}a. The integration range in the Galactic latitude is [$-0${\rlap{$^\circ$}} .41, $0${\rlap{$^\circ$}} .37], while extracting the {\it l}--{\it v} diagram. The dotted yellow line in Figure~\ref{321_twocolor}a separates two velocity components at about $-35.25$ {\kps}. The velocity range and peak velocity of the blue-shifted component are [$-38.25$, $-35.50$] and $-36.75$ {\kps}, respectively, while for the red-shifted component, the velocity range is [$-35.25$, $-27.25$] {\kps}, with a peak velocity of $-32.25$ {\kps}. The spatial distribution of the blue- and red-shifted components is shown using their integrated emission maps in a two-color composite image in Figure~\ref{321_twocolor}b. It is evident from the integrated emission map of the red-shifted component (shown in red in Figure~\ref{321_twocolor}b) that both the HFSs are part of this cloud component. Interestingly, the integrated emission map of the blue-shifted component reveals the presence of a filamentary cloud. Based on the total moment-0 values of the blue-shifted cloud component for the areas bounded by the yellow lines, the total mass of this component is calculated to be about 3.2 $\times 10^2$ {\Msolar}. The red-shifted component shows a spatial fit between its high-intensity region and the low-intensity regions of the blue-shifted filamentary cloud. This feature is referred to as complementary distribution in the literature \citep[see][and references therein]{Maity_2023MNRAS}. To obtain a more compelling demonstration of this complementarity, we moved the red-shifted cloud component in various directions. Based on visual inspection, the best alignment is achieved by shifting the red-shifted component about 4.2 pc to the south-west, as shown in Figure~\ref{321_twocolor}c. The shift estimation is uncertain due to the error in the source distance and our limited knowledge of the exact morphology of the cloud components at the time of collision. The former contributes an uncertainty of 6\% in the shift estimation, while the latter is difficult to quantify. Overall, our estimated shift has a minimum uncertainty of 6\%. 

To better understand the molecular gas distribution toward Hub-1 and Hub-2 in position-velocity ($PV$) space, we extracted the $PV$ diagrams for the arrows A1–A7 shown in Figure~\ref{321_twocolor}b. Arrows A1–A6 are associated with Hub-1, while A7 extends to Hub-2, passing through Hub-1. The width of the slices used to extract the $PV$ diagrams is 3 pixels for A1–A6, and 10 pixels for A7 to cover a larger area. The extracted $PV$ diagrams for the arrows are presented in Figure~\ref{321_PV_diagrams}. The $PV$ diagrams corresponding to A1–A6 show a mixed distribution of two velocity components (see the blue and red arrows). However, for A7, Hub-1 again shows a mixed distribution of two velocity components (see the cyan and red arrows), while other regions, including Hub-2, exhibit only one velocity component (see the white arrows).

\section{Discussion}
\label{G321_Discussion}
We carried out a detailed multi-scale, multi-wavelength observational study of G321. Based on our results, we discuss the role of filaments in mass accumulation and star formation, the massive star-forming activity in G321, and the origin of HFSs.

\subsection{Role of filaments in mass accumulation and star formation in G321}
\label{sec_role_fil_G321}
G321 hosts two HFSs (i.e., HFS-1 and HFS-2; see Figure~\ref{321_filament_maps}), where the hubs (i.e., Hub-1 and Hub-2) are connected with several parsec-scale filaments. As mentioned earlier, filaments are thought to channel molecular gas and dust to their hub, making the hub suitable for MSF \citep{myers_2009,Kumar_2020}. In the case of G321, the total mass accretion rate to Hub-1 through the filaments is {\simi}$2.64\times10^{-2}$ $M_{\odot}$ yr$^{-1}$. For Hub-2, combined mass accretion rate through the filaments is {\simi}$4.17\times10^{-3}$ $M_{\odot}$ yr$^{-1}$. In both cases, the mass accretion rates are comparable to or higher than those observed in other MSFRs, such as {\simi}1 $\times$ 10$^{-3}$ $M_{\odot}$ yr$^{-1}$ for DR 21 ridge \citep{schneider_2010}, 1--3 $\times$ 10$^{-4}$ $M_{\odot}$ yr$^{-1}$ for Serpens \citep{kirk_2013}, 4--7 $\times$ 10$^{-4}$ $M_{\odot}$ yr$^{-1}$ for Monoceros R2 \citep{trevino19}, {\simi}7.40 $\times$ 10$^{-4}$ $M_{\odot}$ yr$^{-1}$ for G310.142+0.758 \citep{Yang_2023ApJ}, {\simi}6.75 $\times$ 10$^{-4}$ $M_{\odot}$ yr$^{-1}$ for G148.24+00.41 \citep{Rawat_2024MNRAS}, and {\simi}1.72 $\times$ 10$^{-3}$ $M_{\odot}$ yr$^{-1}$ for the HFS of RCW 117 \citep{Seshadri_2024MNRAS}. The mass accretion rate for Hub-1 is several times higher than that of Hub-2. 
It is important to note that the potential contribution from filament F4 has not been included in the comparison between the hubs. Moreover, the simple mass accretion model by \citet{kirk_2013} does not apply to F4 because it is connected to both hubs (see Section~\ref{sec_mass_acc_G321}). 
As proposed by \citet{Liu_2019}, the oscillations observed in the average velocity plot along F4 are likely due to its fragmentation process, which is supported by the detection of ATL-5 toward F4. Although no YSO candidate has been detected toward ATL-5, we expect star-forming activity to occur later due to the filament's ongoing fragmentation. The $(M/L)^{\rm Fil}$ values for filaments in G321 varies from about 312 to 1041 {\Msolar} pc$^{-1}$, which are comparable to or higher than the values for other filaments, such as {\simi}385 {\Msolar} pc$^{-1}$ for Orion \citep{Bally_1987ApJ}, {\simi}600 {\Msolar} pc$^{-1}$ for the G11.11 IRDC \citep{Kainulainen_2013AA}, {\simi}115 {\Msolar} pc$^{-1}$ for a subfilament in the G35.39 region \citep{Henshaw_2014MNRAS}, and {\simi}1000 {\Msolar} pc$^{-1}$ for filamentary IRDC 18223 \citep{Beuther_2015AA}. The $(M/L)^{\rm Fil}$ values greatly exceeds the isothermal critical line-mass of about 25 {\Msolar} pc$^{-1}$ for 15 K \citep{Ostriker_1964ApJ}. 
Hence, the turbulent critical line-mass is estimated using, $(M/L)_{\rm crit}^{\rm {tur}}$ $\approx$ 465 $(\sigma^{\rm Fil}_{\rm avg}\,[\rm{km s}^{-1}])^2$ {\Msolar} pc$^{-1}$ \citep{Henshaw_2014MNRAS}. The line-mass ratio (LMR) between ${(M/L)^{\rm Fil}}$ and $(M/L)_{\rm crit}^{\rm {tur}}$ for the filaments is listed in Table~\ref{tab_velograd}. LMR {\simi}1 suggests that turbulence provides major support against the gravitational collapse of the filaments. LMR {\simi}1.3 for F4 implies that it is prone to fragmentation and which is in agreement with our detection of the ongoing filament fragmentation.
 
Hub-1 is more massive than Hub-2, and its higher mass accretion rate suggests a dependence of the mass accretion rate on the hub's mass. Based on this, we propose the relation $\dot{M}_{||} \propto M^{\beta}_{\rm {hub}}$, with $\beta \sim 1.28$ in our case. This finding indicates that the hub's gravitational influence plays a significant role in determining the total mass accretion rate through the filaments. The concept of gas flowing through filaments under the influence of gravity is supported by both observations \citep[e.g.,][]{Williams_2018AA,Wang_2022ApJ,Zhou_2022MNRAS,Zhou_2023MNRAS} and simulations \citep[e.g.,][]{Maity_2024arXiv}.
Although the uncertainties in both $M_{\rm {hub}}$ and $\dot{M}_{||}$ are likely systematic and will not affect the ratio of $M_{\rm {hub}}$ and $\dot{M}_{||}$ between Hub-1 and Hub-2, our estimation of $\beta$ is limited to only two data points. Therefore, the value of $\beta$ could be more precisely constrained with a larger sample of Galactic HFSs. 
The ratio of the total number of YSO candidates observed in VVV data toward Hub-1 to Hub-2 is about 4.5, which closely corresponds to the ratio of their masses. Previously, we found that the mass of the hub influences its total mass accretion rate. Thus, this study highlights a potential connection between the mass of the hub, its mass accretion rate, and its star-forming activity.
The total mass of Hub-1 and Hub-2 is about 2 $\times$ 10$^4$ {\Msolar}, which is about 25\% of the total mass of the molecular cloud. The combined mass accretion rate for Hub-1 and Hub-2 is about $3.1\times10^{-2}$ $M_{\odot}$ yr$^{-1}$. Given that the mass accretion rate increases with the mass of the hub, they are expected to accumulate an additional 25\% of the cloud's total mass in less than 0.6 Myr unless disrupted by the feedback from the newly formed massive stars. This underscores the significant role of filaments in mass accumulation and in creating high-density regions that are conducive to MSF. 

Interestingly, we have detected several small-scale ($\lesssim$ 1 pc) filament candidates within Hub-1 and Hub-2, indicated by black arrows (see Figure~\ref{321_filament_maps}c). As proposed by \citet[][]{Dewangan_2024MNRAS}, the presence of such small-scale filaments can be further confirmed through James Webb Space Telescope ({\it JWST}) NIR images in absorption. The detection of small-scale filaments, along with large-scale ($>1$ pc) filaments connected to hubs, is important for assessing the self-similar hierarchical HFS scenario introduced by \citet{Zhou_2022MNRAS}, which proposes filamentary mass accretion occurring across multiple scales. Therefore, {\it JWST} NIR observations of Hub-1 and Hub-2 will be valuable for further investigation.

\subsection{The massive star-forming activity in G321}
\label{sec_MSF_G321}
The presence of four compact {\htwo} regions, two HFSs with high mass accretion rates (i.e., $> 10^{-3}$ $M_{\odot}$ yr$^{-1}$) to their hubs, along with evidence of filament fragmentation in F4, makes this region particularly intriguing for studying its potential for MSF. As shown in Figure~\ref{321_ATLASGAL_154}, Hub-1 hosts ATB-1, which further fragments into ATL-3 and ATL-4. ATB-1 and Leaf-3 satisfy mKP-10 condition for MSF \citep{Kauffmann_2010ApJ}. Therefore, HFS-1 is actively engaged in MSF, which is further supported by the detection of three compact {\htwo} regions within Hub-1. By comparing the $N_{\rm uv}$ values for these {\htwo} regions (see Table~\ref{tab1_htwo}) with the theoretical predictions from \citet{Panagia_1973}, we determined that the ionizing sources are massive stars of spectral type B1V–B3V. Furthermore, we estimated the minimum stellar mass for a massive star that could be expected from the ATL-3 using the formula from \citet[][]{Sanhueza_2017ApJ}:
\begin{equation}
M_{\rm exp} = \left(\frac{0.3}{\epsilon_{\rm sfe}}\frac{17.3}{M_{\rm clump}} + 1.5\times 10^{-3}\right)^{-0.77}~,
\label{eqn-IMF-m-max-ap}
\end{equation}
where ${\epsilon_{\rm sfe}}$ is the star formation efficiency. This formula was derived assuming the range of stellar masses between 0.08 and 150 {\Msolar} and adopting the initial mass function from \citet{Kroupa2001MNRAS}. The study by \citet{Wells_2022MNRAS} determined that ${\epsilon_{\rm sfe}}$ = 0.2--0.3 for a clump of a few hundred solar masses. Hence, using ${\epsilon_{\rm sfe}}$ = 0.2--0.3, and $M_{\rm clump}$ = 749 {\Msolar} for ATL-3 in Equation~\ref{eqn-IMF-m-max-ap}, we found $M_{\rm exp}$ {\simi}13--17 {\Msolar}. Given the high mass accretion rate of Hub-1, this can be considered a lower limit for the maximum stellar mass within Hub-1, and we may expect a small cluster of O-type stars to form in Hub-1.

We have estimated the thermal Jeans mass for ATL-3 to be about 3 {\Msolar} using Equation~6 from the study of \citet{Palau_2015MNRAS}.
This estimate is in agreement with the masses of the ALMA leaves/cores detected within ATL-3, which range from 1--9 {\Msolar}, with an average of about 4 {\Msolar} (see Table~\ref{tab_alma_tree}). Subsonic/transonic turbulence is sufficient to explain the existence of the ALMA leaves above the Jeans mass (i.e., $> 3$ {\Msolar}). However, the thermal Jeans mass estimated for the ALMA branches, which is in the range of 0.1--0.2 {\Msolar}, does not match the masses of their substructures (i.e., further branches and leaves). Therefore, the ALMA branches may represent ensembles of their substructures that evolve as the substructures themselves evolve, not through step-by-step thermal fragmentation, but rather through a simultaneous formation process as discussed in \citet{Morii_2024ApJ}. It is important to note that the lack of molecular line data compatible with the resolution of the ALMA Band-7 continuum image prevents us from conducting any analysis related to turbulence at the core-scale. 

Although ATL-4 (in Hub-1), ATL-5 (toward Filament F4), and ATL-6 (in Hub-2) marginally satisfy mKP-10 condition for MSF, the large errors (at least {\simi}20\%; see Section~\ref{dendrogram_ATLASGAL}) associated with the mass estimation cast doubt on their potential for MSF. Additionally, the absence of radio continuum emission from these leaves  suggests that they are not currently active in MSF. However, since ATL-4 and ATL-6 are part of Hub-1 and Hub-2, respectively, and both actively accumulating material through the filaments, these leaves may evolve into potential MSF sites. Similarly, ATL-5, being part of a fragmenting filament, has the potential to become active in MSF at a later stage. The presence of a compact {\htwo} region (driven by a B1V type source) toward ATL-2 clearly indicates ongoing massive star-forming activity. Interestingly, ATL-2 falls below the mKP-10 limit for MSF, a result that aligns with approximately 10\% of the MSF sites listed in \citet{Beuther_2002ApJ}, as noted by \citet{Kauffmann_2010ApJ}. This suggests that the mKP-10 limit serves as an approximate threshold for MSF, and follow-up studies are required for such sources, including ATL-2. 

According to  \citet{motte_2018}, MSF begins in massive dense cores (MDCs) that contain low-mass pre-stellar cores. Over time, these MDCs develop into infrared (IR)-quiet protostellar objects that harbor low-mass stellar embryos ($<$ 8\,{\Msolar}). These embryos then evolve into IR-bright high-mass protostars ($>$ 8\,{\Msolar}) through gravitational inflows, eventually maturing into massive stars capable of creating {\htwo} regions. 
In the case of Hub-1, all the structures from ATB-1 and ATL-3 to the ALMA leaves/cores satisfy the mKP-10 condition for MSF. Interestingly, two ALMA branches (IDs: 4 and 13) have masses above 20\,{\Msolar}; however, they are further fragmented into low-mass cores. The detection of YSO candidates, along with 22 GHz H$_2$O and Class~I 95 GHz CH$_3$OH maser emissions, indicates ongoing star-forming activity in some of the ALMA low-mass cores. Thus, the presence of only low-mass cores actively forming stars within a hub that satisfies the condition for MSF supports the applicability of the idea proposed by \citet{motte_2018}. 
It is important to note that the ALMA data cover a small area of ATL-3, which is located at the edge of the {\htwo}-2 region (not shown in the figure). As a result, it appears that the main driving source of the {\htwo}-2 region is beyond the coverage of the ALMA observations. Therefore, a high-resolution ALMA observation covering the entire area of ATB-1 would be beneficial for revealing the complete core mass function, including the cores associated with the {\htwo} regions.
The absence of {\htwo} regions toward Hub-2 indicates that HFS-2 can be in the early stages of Motte's scheme. However, the presence of {\htwo} regions toward Hub-1 and the ATL-2 suggests that they are in the later stages of evolution. Hence, we observe significant differences in the evolutionary stages of MSF at our target site, G321.

\subsection{Origin of the hub-filament systems}
\label{sec_CCC_G321}
Despite being part of the same cloud component (see Section~\ref{sec_mol_G321}), HFS-1 and HFS-2 show significant differences in terms of the mass of their hubs, mass accretion rate through the filaments, number of color excess sources present in the hubs, and their potential for MSF. In this section, we attempt to explain  the origin of these HFSs.
As earlier mentioned, a convincing picture of the complementary spatial distribution between the blue-shifted filamentary cloud component and the red-shifted component was achieved by shifting the red-shifted component toward the south-west direction (see Figure~\ref{321_twocolor}c). The complementary distribution of two cloud components with a spatial shift in the plane of sky is a well-recognized signature of CCC \citep[e.g.,][]{Fukui_2017PASJ,Dewangan2018,fukui21,tsuge21exg,Maity_2023MNRAS}. The colliding molecular clouds create a compressed layer at their interface with an intermediate velocity, connecting the velocities of the parent molecular cloud components in the $PV$ diagram, which is referred to as a bridge feature \citep[e.g.,][]{fukui_2014,haworth_2015a,haworth_2015b,torri_2017,Sano_2018PASJ,fukui21,Priestley2021,maity_W31}. It is worth noting that the $PV$ diagram is highly dependent on the morphology and density of the colliding cloud components. The angle ($\theta_\mathrm{col}$) between the observer's line of sight and the axis of collision also influences the $PV$ diagrams \citep{fukuia_2018}. In the worst-case scenario, i.e., at $\theta_\mathrm{col}$ = 90$^{\circ}$, the colliding clouds become indiscernible in velocity space \citep[e.g.,][]{takahira_2014,Priestley2021}.  

The intriguing feature of the {\it l}--{\it v} diagram for G321 is the presence of two cloud components and the absence of the bridge feature. Several studies based on numerical simulations demonstrated the temporal evolution of $PV$ diagrams in the context of CCC \citep[][]{takahira_2014,haworth_2015b,Maity_2024arXiv}. \citet{haworth_2015b} argued that the bridge feature may disappear if one cloud component punches through the other. In such cases, the compressed layer will mix with the cloud component passing through the other \citep{Maity_2024arXiv} in the velocity space. The observed morphology of the blue- and red-shifted cloud components toward HFS-1 is consistent with the scenario in which one cloud component punches through the other. Specifically, the red-shifted component appears to have passed through the blue-shifted filamentary cloud, creating a cavity or low-intensity region in the filament. As a result, this scenario does not align with the proposal by \citet{Kumar_2020}, where HFSs form through the merging or overlapping of filaments. In addition, the detection of the mixing of two velocity components in the $PV$ diagrams along A1--A7 toward Hub-1 (see Figure~\ref{321_PV_diagrams}) suggests that the compressed layer created during the collision has merged with the red-shifted cloud component in velocity space. Therefore, this study suggests that the formation of HFS-1 was possibly triggered by the collision of the red-shifted cloud component with the blue-shifted filamentary cloud. The collision timescale can be estimated using the formula, $t_{\mathrm{collision}} = {l_{\mathrm{loc}}}/{v_{\mathrm{loc}}}$ \citep{Maity_2023MNRAS}. In this expression, $l_{\mathrm{loc}}$ and $v_{\mathrm{loc}}$ denote the spatial shift along the collision axis and the intrinsic collision velocity, respectively. These quantities are related to the observed spatial shift ($l_{\mathrm{obs}}$) and observed collision velocity ($v_{\mathrm{obs}}$) by the relations $l_{\mathrm{loc}} = {l_{\mathrm{obs}}}/{\sin\theta_\mathrm{col}}$ and $v_{\mathrm{loc}} =  {v_{\mathrm{obs}}}/{\cos\theta_\mathrm{col}}$. The detection of two cloud components in the {\it l}--{\it v} diagram for our target sites, along with the necessity of spatial shift for complementary distribution, indicates that $\theta_\mathrm{col}$ is significant and less than 90$^{\circ}$. Hence, we assume $\theta_\mathrm{col}$ to be within the range of 30$^{\circ}$ to 60$^{\circ}$. Using $l_{\mathrm{obs}}$ = 4.2 pc, $v_{\mathrm{obs}}$ = 4.5 {\kps}, and $\theta_\mathrm{col}$ = 30$^{\circ}$, 45$^{\circ}$, and 60$^{\circ}$, the corresponding collision timescales ($t_{\mathrm{collision}}$) are estimated to be 1.58, 0.91, and 0.53 Myr, respectively. 
These timescales are consistent with the dynamical timescales of the {\htwo} regions (i.e., $t_{\rm {dyn}} <$ 0.25 Myr) and the mean age of the {\cone}, which is about 0.44 Myr \citep{evans_2009}. This study, therefore, suggests that a CCC event approximately 1 Myr ago triggered star-forming activity in the region of the collision (i.e., toward Hub-1). The collision also contributed to an increase in the total mass and density of the gas in the collision region, leading to its high mass accretion rate and massive star-forming activity. 
In the case of CCC, the collision velocity has a significant influence on MSF, which can be understood through a direct comparison between G321/Hub-1 and a few CCC sites from literature such as NGC 6334, W31 complex, and AFGL 5180 \citep[][]{CCC3,maity_W31,Maity_2023MNRAS}. The observed collision velocity for AFGL 5180, G321/Hub-1, W31 complex, and NGC 6334 are about 3.7 {\kps} \citep[][]{Maity_2023MNRAS}, 4.5 {\kps}, 8 {\kps} \citep[][]{maity_W31}, and 12 {\kps} \citep[][]{CCC3}, respectively. Although AFGL 5180, G321/Hub-1, and the W31 complex have a similar collision timescale of about 1 Myr, they show variations in massive star-forming activity due to their distinct collision velocities. AFGL 5180 contains one hyper-compact {\htwo} region ($\lesssim$ 0.05 pc), while G321/Hub-1 has three compact {\htwo} regions associated with B1V--B3V stars. On the other hand, the W31 complex hosts a classical {\htwo} region powered by several O-type stars. It suggests higher collision velocities lead to the creation of more massive stars in terms of both their number and mass. This aligns with the findings of \citet{fukui21} (see Figures 10 and 11 in their work), which demonstrated that increased collision velocities result in higher column densities, facilitating the formation of large numbers of massive stars. Additionally, the formation of several O-type stars in NGC 6334 within a relatively shorter timescale  (i.e., $t_{\mathrm{collision}}$ {\simi}0.1 Myr) shows that massive star-forming activity accelerates with increasing collision velocity.

The filaments connected to Hub-2 (i.e., F4, F5, and F6) have a relative velocity of about 1.5 {\kps} as shown in the two-color composite image in Figure~\ref{2color_hub2_leaf2}a. Interestingly, we detected at least three filaments toward ATL-2 through visual inspection of the moment-0 maps of {\tco} emission for the velocity ranges [$-32.25,-31.75$] and [$-29.25,-28.75$] {\kps}. These filaments are marked using dashed-dotted lines over the two-color composite image in Figure~\ref{2color_hub2_leaf2}b. These filaments were not visible in Figure~\ref{321_filament_maps} due to their weak emission strength in the integrated intensity map for the wide velocity range (i.e., [$-38.25,-27.25$] {\kps}). ATL-2 is located precisely at the junction of these filaments, making the entire configuration a candidate HFS (C-HFS). 
The differing velocities of the filaments for the HFS-2 and C-HFS support the idea proposed by \citet{Kumar_2020}, where HFSs form through the merging or overlapping of filaments. Further insights into this candidate HFS can be gained through molecular line observations with improved sensitivity and resolution.

In their recent study, \citet{Semadeni_2024arXiv} proposed that CCC and the merging/overlapping of filaments fall under the gravity-driven global hierarchical collapse \citep[GHC; e.g.,][]{vazquez_2009, vazquez_2017, vazquez_2019} scenario. However, as shown by \citet[][]{inoue_2013,inoue18}, filaments can form in the absence of gravity, where turbulence and shock compression due to collision play a major role. Specifically, collisions lead to the rapid compression of turbulence-driven inhomogeneous structures into filaments. Therefore, both turbulence and collision have a distinct role in the formation of filaments. Finally, due to the effect of gravity, the filaments make a common junction to form HFSs \cite[see][]{Maity_2024arXiv}. Similarly, the merging of filaments amplifies the mass and density in the overlapping zones, enhancing the gravitational potential and driving gas flow along the filaments \cite[see][]{Kumar_2020}. Therefore, gravity plays an evident role in the later stages of evolution for both scenarios (i.e., CCC and the merging/overlapping of filaments). However, the distinct effects of collisions and merging cannot be ignored.

\section{Summary and Conclusions}
\label{sec_sum_JP}
To understand the formation mechanisms of HFSs, mass accretion through the filaments, and MSF, we conducted a multi-scale and multi-wavelength observational investigation of G321.  
The key findings of this study are summarized below.

\begin{enumerate}
\item G321 hosts multiple HFSs: HFS-1, HFS-2, and a C-HFS. HFS-1 and HFS-2 are connected by a filament (i.e., F4), which shows clear signs of fragmentation. This is a unique result, considering that such complex sites are rare in the literature.

\item HFS-1 and HFS-2 exhibit high mass accretion rates (i.e., $\dot{M}_{||}$ $> 10^{-3}$ $M_{\odot}$ yr$^{-1}$). We found $\dot{M}_{||} \propto M^{\beta}_{\rm {hub}}$ with $\beta \sim 1.28$, suggesting the gravitational influence of the hub on mass accretion through filaments. The study of color excess sources demonstrates a possible connection among the mass of the hub, its mass accretion rate, and its star-forming activity. 
 
\item Considering the presence of three compact {\htwo} regions driven by B1V--B3V type stars in Hub-1 and its high $\dot{M}_{||}$ values, we may expect a small cluster of O-type stars to form in Hub-1.
  
\item The analysis of ALMA Band-7 continuum data inside Hub-1/ATL-3 revealed several high-mass branches ($M$ {\simi}8--33\,{\Msolar}), which are further fragmented into branches and leaves. We have calculated the mass of these leaves/cores to be about 1--9\,{\Msolar}, which is in agreement with the thermal Jeans mass of ATL-3 with subsonic/transonic turbulence.

\item A significant difference in the evolutionary stages of MSF is observed in G321. The absence of {\htwo} regions toward Hub-2 indicates that HFS-2 can be in the early stages of Motte's scheme \citep{motte_2018}. However, the presence of compact {\htwo} regions toward Hub-1 and the ATL-2 suggests that they are in the later stages of evolution.
 
\item HFS-1 shows a clear signature of collision with a filamentary cloud about 1 Myr ago, suggesting that the CCC may have triggered the formation of HFS-1. In the cases of HFS-2 and C-HFS, their constituent filaments exhibit relative velocities ($\gtrsim$1 {\kps}), indicating possible formation through the merging or overlapping of filaments.

\end{enumerate}
Overall, we have found G321 to be an interesting MSF region, hosting several HFSs. This study clearly shows that interacting filaments (either by collision or merging/overlapping) can form high-density regions susceptible to MSF. It is important to note that we have not explored the formation mechanisms of the individual filaments in this study, which can be addressed in future work.

%
\begin{figure*}
\centering
\includegraphics[width= 0.50 \textwidth]{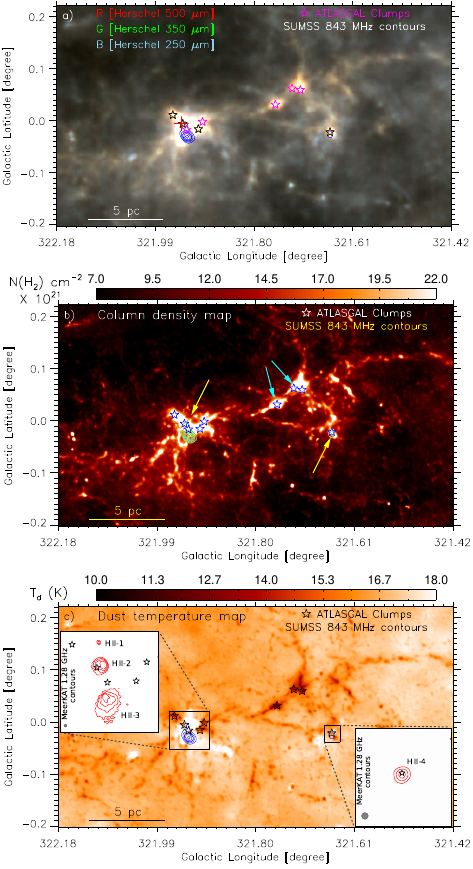}
\caption{(a) {\it Herschel} three color-composite image. The colors red, green, and blue present  {\it Herschel} 500, 350, and 250 {\micro} continuum images, respectively. The plus symbol indicates the position of the 22 GHz H$_2$O and Class~I 95 GHz CH$_3$OH maser emissions. The asterisks denote the positions of the ATLASGAL clumps from \citet{Atlasgal}. Notably, the clumps exhibiting active outflows are marked in black \citep{Yang2022}.  
Panels (b) and (c) display {\it Herschel} column density (i.e., $N$(H$_2$)) and dust temperature (i.e., $T_{\rm d}$) maps, respectively. The contours in each panel show the SUMSS 843 MHz radio continuum emission. The contour levels are at about [3, 6, 9, 12, and 15] $\times$ $\sigma$, where 1$\sigma$ {\simi}1 mJy beam$^{-1}$. In panel ``b,'' arrows highlight high column density regions. Yellow arrows indicate areas characterized by higher dust temperatures and the detection of radio emissions, while cyan arrows denote regions with lower dust temperatures and no detectable radio sources. The rectangular regions highlighted in panel ``c" are zoomed-in within the insets. The contours in the top-left and bottom-right insets represent MeerKAT 1.28 GHz radio continuum emission at [5, 15, 30, and 50] $\times$ $\sigma$ and [5, 50, and 160] $\times$ $\sigma$, respectively, where 1$\sigma$ {\simi}20 {$\mu$}Jy beam$^{-1}$. A gray circle at the bottom-left corner of each inset indicates the beam size ($\sim$8{\as}$\times$8{\as}) of the MeerKAT data. A scale bar of 5 pc is shown in each panel at a distance of 1.98 kpc.}
\label{G321_herschel_maps}
\end{figure*}

\begin{figure*}
\centering
\includegraphics[width= \textwidth]{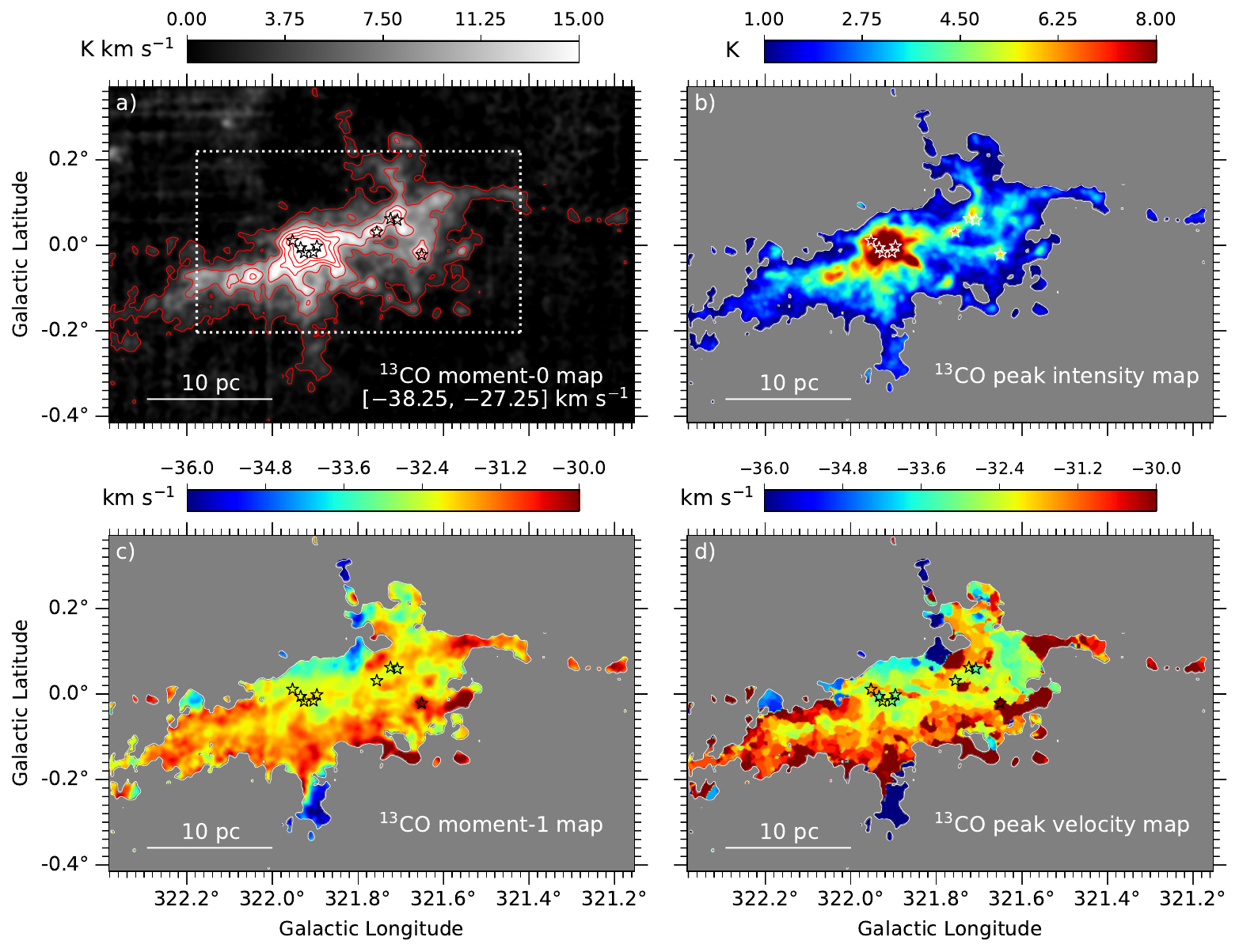}
\caption{(a) SEDIGISM {\tco} integrated intensity (i.e., the moment-0) map at $v_{\mathrm {lsr}}$ of [$-$38.25, $-$27.25] {\kps}.
The contour levels are at [3, 11, 19, 27, 35, and 43]$\times \sigma$, where 1{$\sigma$} {\simi}0.6 K {\kps}. The dotted white rectangle indicates the area of {\it Herschel} continuum images shown in Figure~\ref{G321_herschel_maps}. (b) The peak intensity map for the same molecular line data. (c) The intensity-weighted velocity (i.e., the moment-1) map. (d) The velocity distribution corresponding to the peak intensity, referred to here as the `peak velocity map.' The images in panels ``b,'' ``c,'' and ``d'' are clipped below 3$\sigma$ level of the moment-0 map. The asterisks are the same as Figure~\ref{G321_herschel_maps}. A scale bar of 10 pc is shown in each panel.}
\label{321_molecular_maps}
\end{figure*}

\begin{figure*}
\centering
\includegraphics[width= \textwidth]{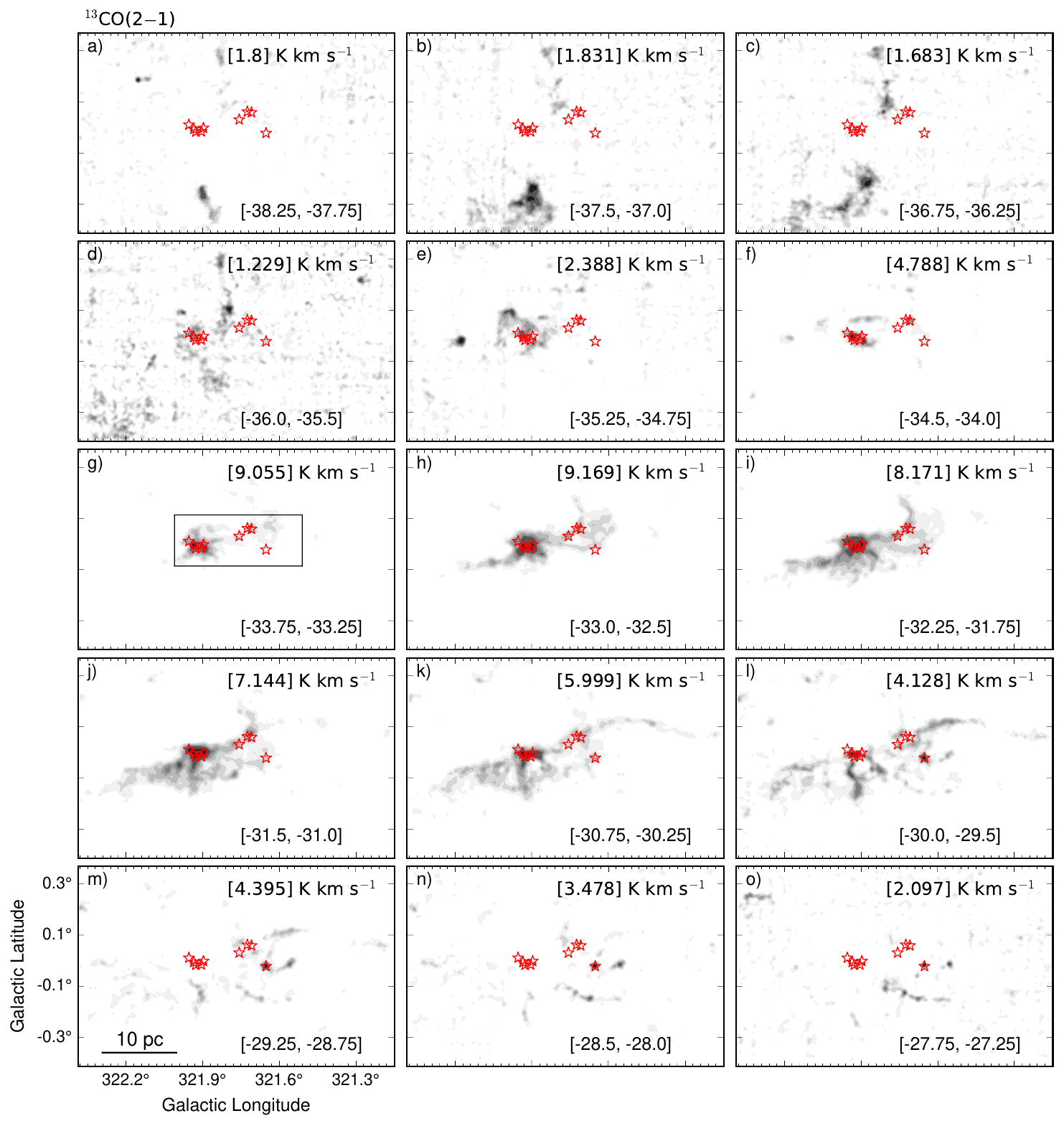}
\caption{
Panels (a)--(o) present integrated intensity maps of {\tco} emission (using filled contours) for $v_{\mathrm {lsr}}$ starting from $-$38.25 to $-$27.25 {\kps} with an interval of about 0.75 {\kps}. The contour levels are at [0.1, 0.2, 0.3, 0.4, 0.5, 0.6, 0.8, and 1.0] $\times$ peak moment-0 values, which are mentioned in the respective panels. The black rectangle shown in panel ``g'' is zoomed-in using the ATLASGAL 870 {\micro} continuum image in Figure~{\ref{321_ATLASGAL_154}a}. A scale bar of 10 pc is shown in panel ``m.''}
\label{321_channel_maps}
\end{figure*}

\begin{figure*}
\centering
\includegraphics[width= 0.5\textwidth]{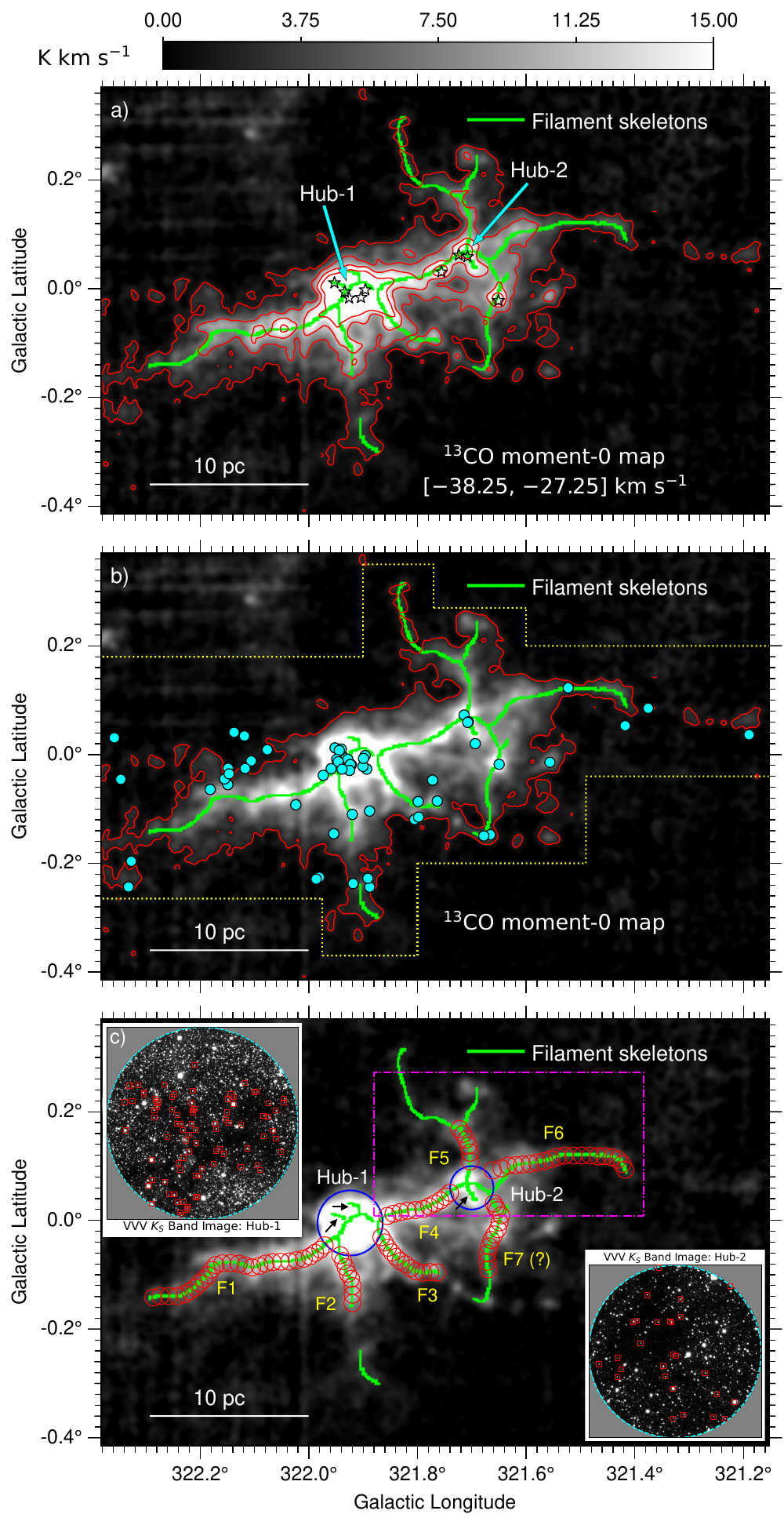}
\caption{(a) The {\it getsf}-identified filament skeletons are marked over the moment-0 map of {\tco} emission. The moment-0 map is identical to Figure~\ref{321_molecular_maps}a, and the contour is at 3$\sigma$ level. (b) The distribution of Class I YSO candidates is shown in the region bounded by dotted yellow lines, with solid cyan circles outlined in black. (c) Two blue circles over the moment-0 map indicate the possible extent of Hub-1 and Hub-2.
Black arrows indicate small-scale ($\lesssim 1$ pc) filaments within the hubs. Both the hubs are zoomed-in using VVV $K_\mathrm{s}$ band image in the insets. The red rectangles in the insets indicate the embedded color excess sources with $H - K_{\mathrm s} > 1.8$. The red circular regions over the filament skeletons are utilized to extract the average velocity and velocity dispersion along the filaments, which are shown in Figure~\ref{321_velocity_gradient}. The area of the magenta dashed-dotted rectangular region is shown using a two-color composite image in Figure~\ref{2color_hub2_leaf2}a.}
\label{321_filament_maps}
\end{figure*}

\begin{figure*}
\centering
\includegraphics[width= \textwidth]{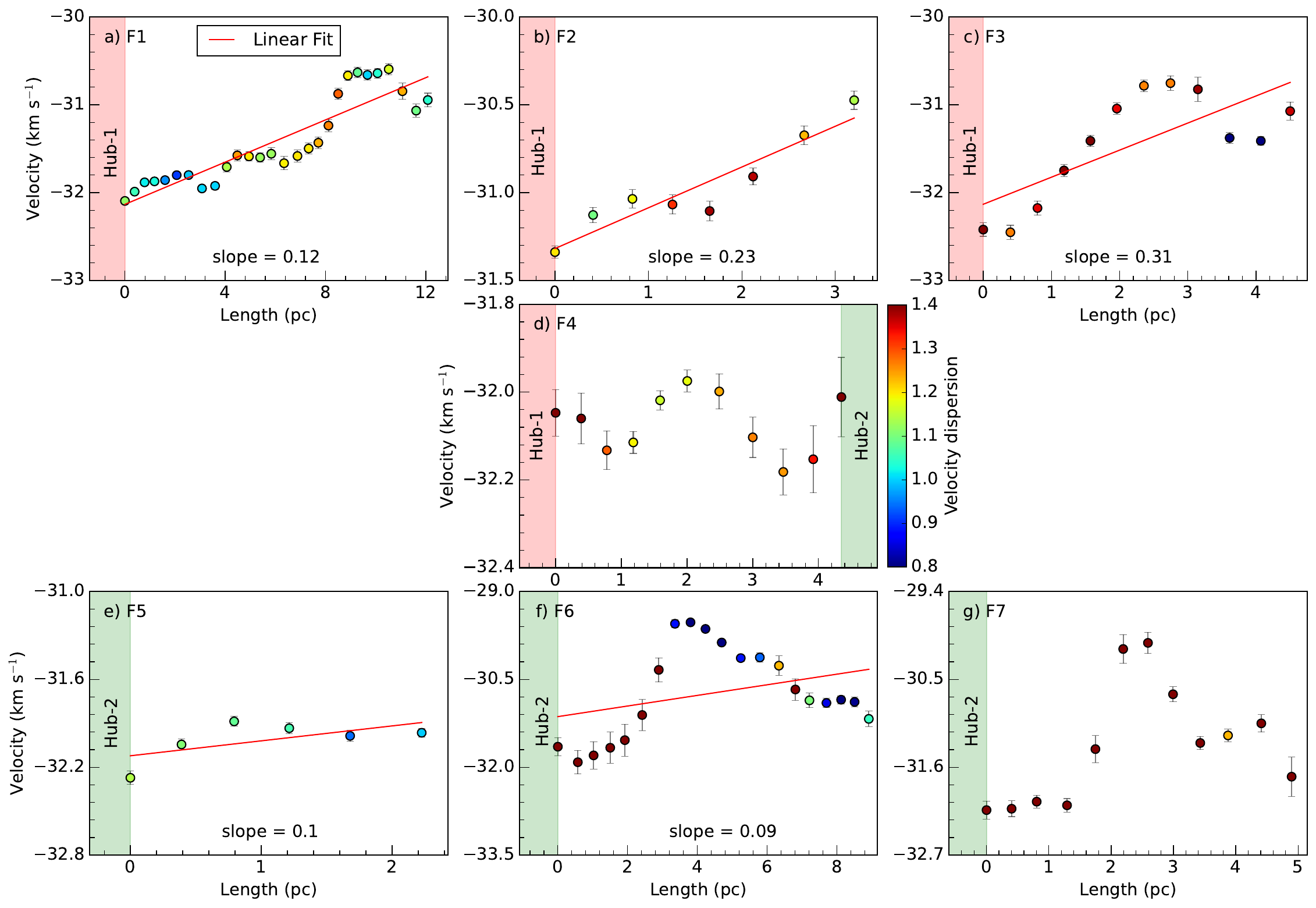}
\caption{Panels (a)–(g) show the average velocities along the filaments F1--F7 for the circular regions depicted in Figure~\ref{321_filament_maps}c. The velocity dispersion for these regions is shown using the color scale. The red straight lines indicate the best linear fit to the average velocity distributions. The slopes of the best-fit lines for filaments F1, F2, F3, F5, and F6 are provided in their respective panels.}
\label{321_velocity_gradient}
\end{figure*}

\begin{figure*}
\centering
\includegraphics[width= \textwidth]{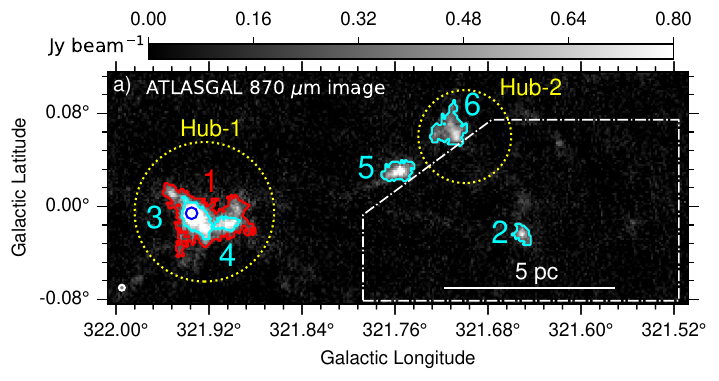}
\includegraphics[width= 0.5\textwidth]{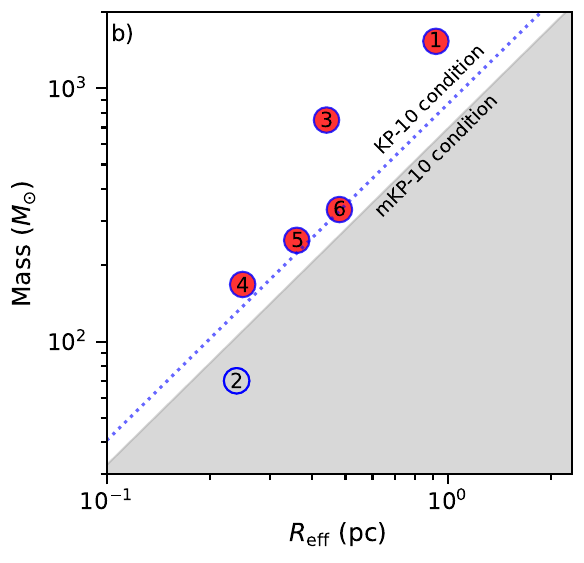}
\caption{(a) The panel shows {\it astrodendro}-identified hiererchical structures in ATLASGAL 870 {\micro} continuum image. The branch is highlighted with a red contour, while the leaves are presented in cyan. The yellow dotted circles indicate the extent of the hubs (i.e., Hub-1 and Hub-2, as shown in Figure~\ref{321_filament_maps}c). The blue circle inside Leaf-3 is further zoomed-in using the ALMA Band-7 continuum image in Figure~\ref{321_ALMA_images}a. The area of the white polygon is shown using a two-color composite image in Figure~\ref{2color_hub2_leaf2}b. The white circle at the bottom-left corner of the image presents the beam size ({\simi}18{\rlap {\as}} .2 $\times$ 18{\rlap {\as}} .2) of the ATLASGAL data. A scale bar of 5 pc is added to this panel. (b) The mass--effective radius plot of the {\it astrodendro}-identified structural components. The blue dotted line presents the KP-10 condition for MSF \citep{Kauffmann_2010ApJ}. The white region above the gray shaded area corresponds to the mKP-10 condition for MSF, and the structural components that satisfy this condition are shaded in red.}
\label{321_ATLASGAL_154}
\end{figure*}

\begin{figure*}
\centering
\includegraphics[width= 0.48\textwidth]{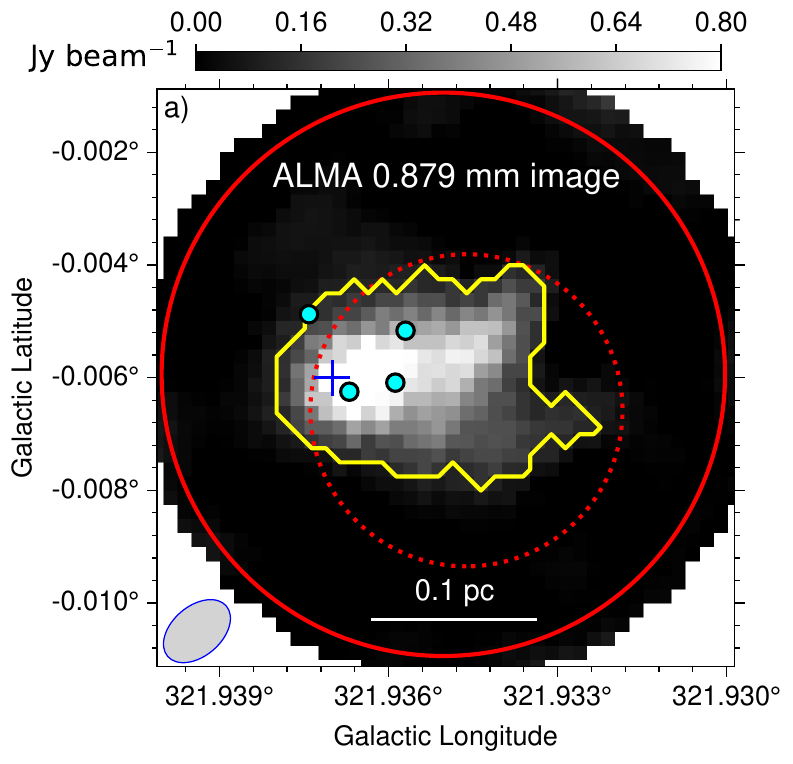}
\includegraphics[width= 0.48\textwidth]{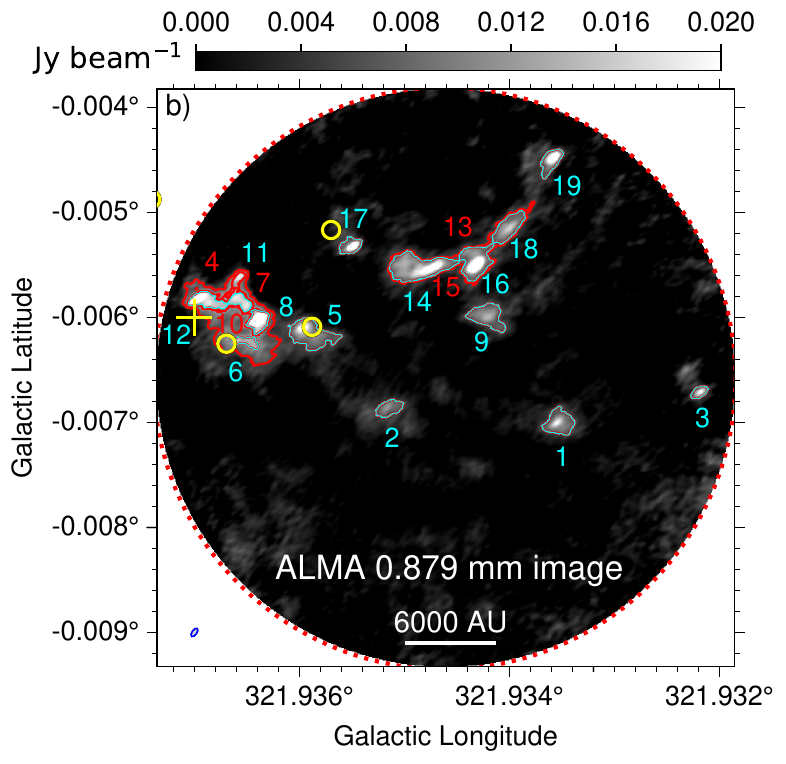}
\includegraphics[width= 0.59\textwidth]{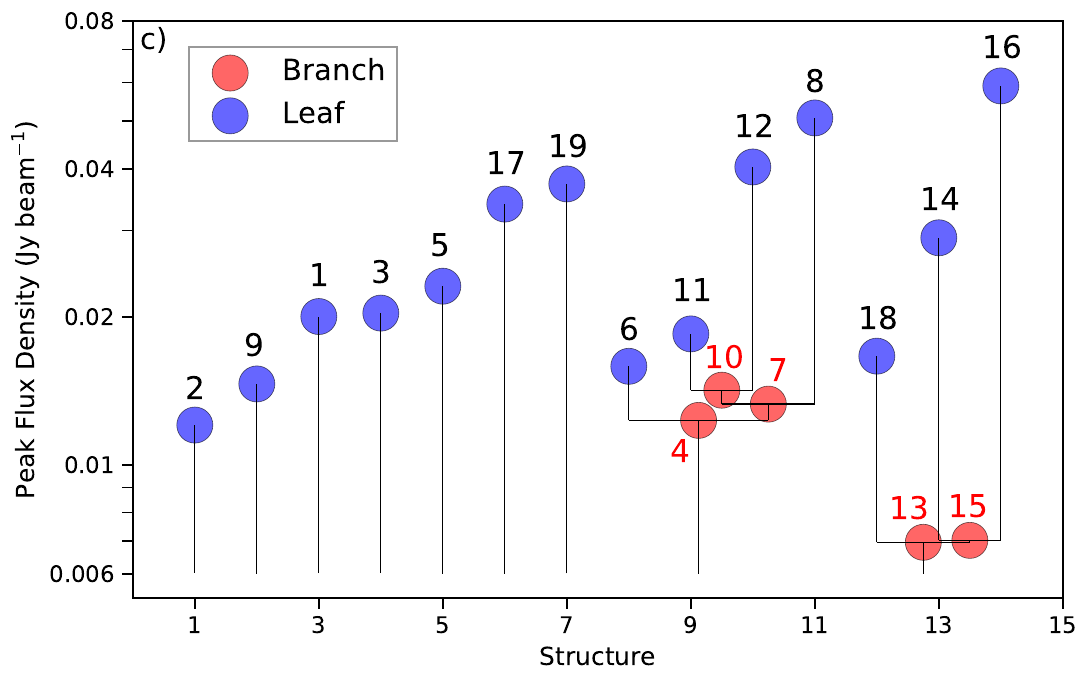}
\includegraphics[width= 0.40\textwidth]{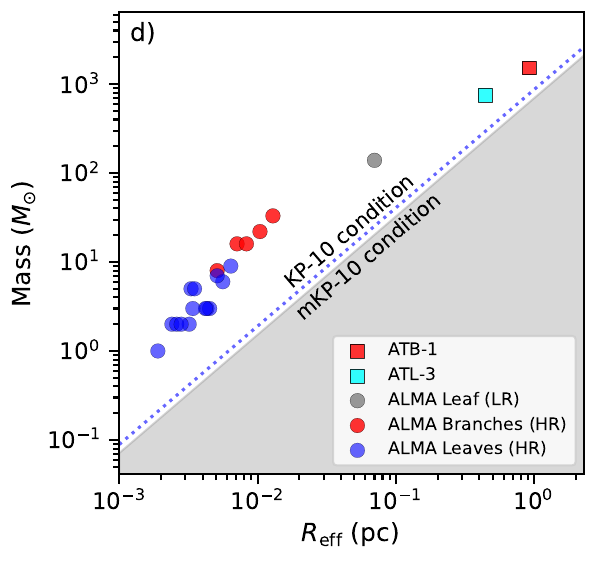}

\caption{
(a) ALMA Band-7 continuum image (beam size $\sim$4\rlap.{$''$}9$\times$3\rlap.{$''$}1) of the area highlighted in Figure~\ref{321_ATLASGAL_154}a obtained with ALMA 7 m array. The solid red circle represents the extent of the blue circle in Figure~\ref{321_ATLASGAL_154}a.
The yellow contour presents the {\it astrodendro}-identified structure. A scale bar of 0.1 pc is marked in the panel. The area under the dotted red circle is further zoomed-in in panel ``b.'' (b) Zoomed-in view of the target area using ALMA Band-7 continuum image of beam size $\sim$0\rlap.{$''$}33$\times$0\rlap.{$''$}16 obtained with ALMA 12 m array. The {\it astrodendro}-identified structures, i.e., the branches and leaves are marked in colors red and cyan, respectively. A scale bar of 6000 AU is shown on the panel. In panels ``a'' and ``b'', the plus symbol indicates the position of the 22 GHz H$_2$O and Class I 95 GHz CH$_3$OH maser emissions, and the circles represent Class I YSO candidates. The ellipses in the bottom-left corners of panels ``a'' and ``b'' indicate the beam size of the data. (c) The {\it astrodendro}-identified dendrogram tree for the structural components shown in panel ``b.'' (d) The mass--effective radius plot of the {\it astrodendro}-identified structural components from ATB-1 to ALMA leaves/cores. For ALMA, ``LR'' and ``HR'' stand for low- and high-resolution data. The blue dotted line and the gray 
shaded area are identical to Figure~\ref{321_ATLASGAL_154}b.}
\label{321_ALMA_images}
\end{figure*}

\begin{figure*}
\centering
\includegraphics[width= 0.50\textwidth]{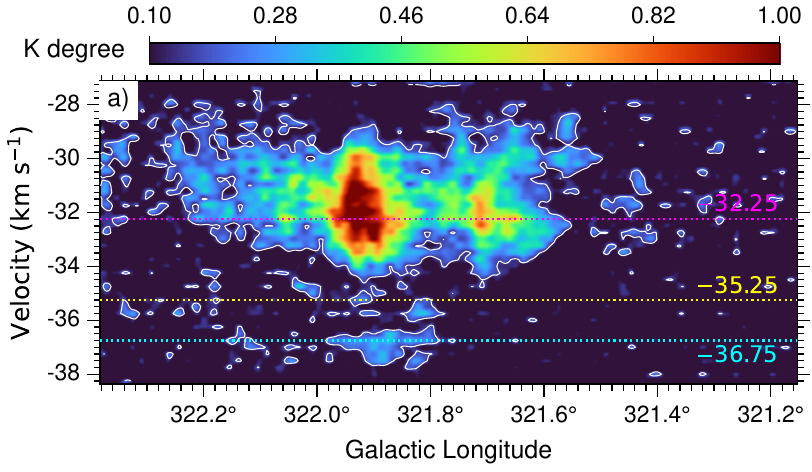}
\includegraphics[width= 0.50\textwidth]{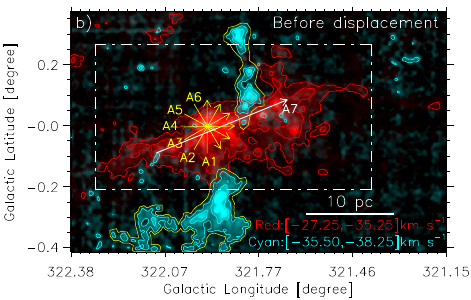}
\includegraphics[width= 0.50\textwidth]{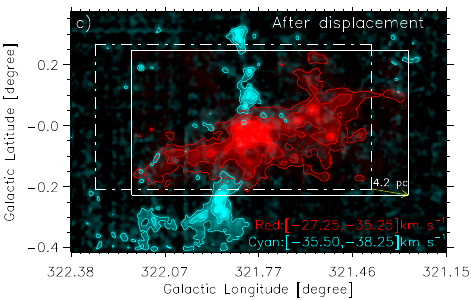}
\caption{(a) The Galactic longitude-velocity (i.e., ${l}$--$v$) diagram for {\tco} data. The contour is at 10$\sigma$, where 1$\sigma$ {\simi}0.016 K degree. The integration range for the Galactic latitude is [$-$0.41, 0.37]\,degree.
The yellow dotted line separates two velocity components at about $-35.25$ {\kps}. The peak velocities for the blue- and red-shifted components are shown with cyan and magenta dotted lines, respectively. (b) The spatial distribution of the blue- and red-shifted components. The blue- and red-shifted components are integrated for the velocity ranges [$-38.25$, $-35.50$] and [$-35.25$, $-27.25$] {\kps}, respectively. The contour levels for the blue- and red-shifted components are at [3$\sigma$, 6$\sigma$, 9$\sigma$] and [6$\sigma$, 15$\sigma$, 30$\sigma$], respectively. The 1$\sigma$ values are 0.3 and 0.5 K {\kps} for integrated intensity maps of the blue- and red-shifted components, respectively. The areas bounded by the yellow lines are considered for calculating the total mass of the blue-shifted component. The position-velocity ({\it PV}) diagrams are extracted along the arrows highlighted in the panel ``b'', which are shown in Figure~\ref{321_PV_diagrams}. (c) Same as panel ``b'', with red-shifted component shifted by about 4.2 pc. The dashed-dotted and solid white rectangles present the initial and final positions of the red-shifted cloud component, respectively.}
\label{321_twocolor}
\end{figure*}

\begin{figure*}
\centering
\includegraphics[width= 0.7\textwidth]{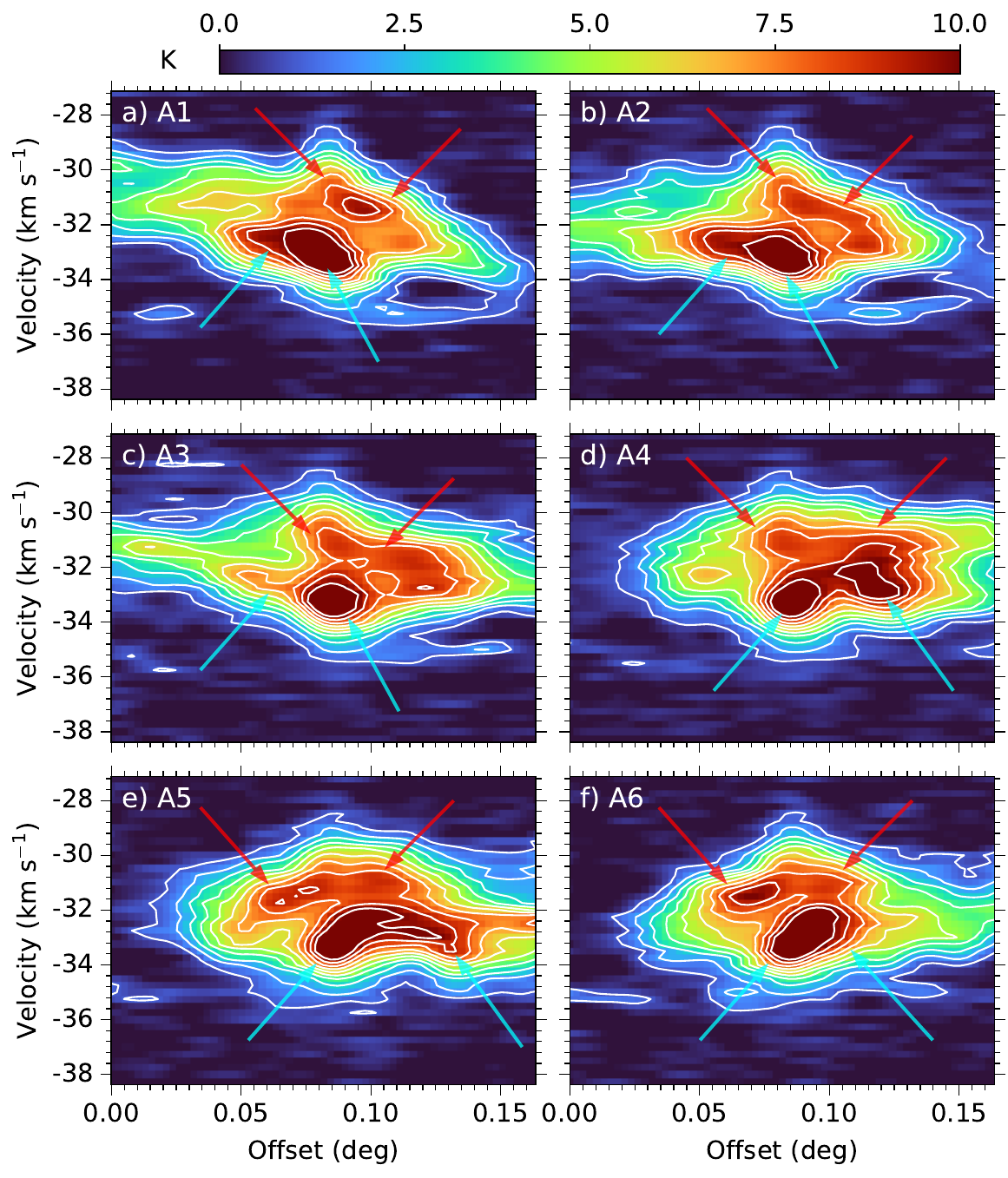}
\includegraphics[width= 0.7\textwidth]{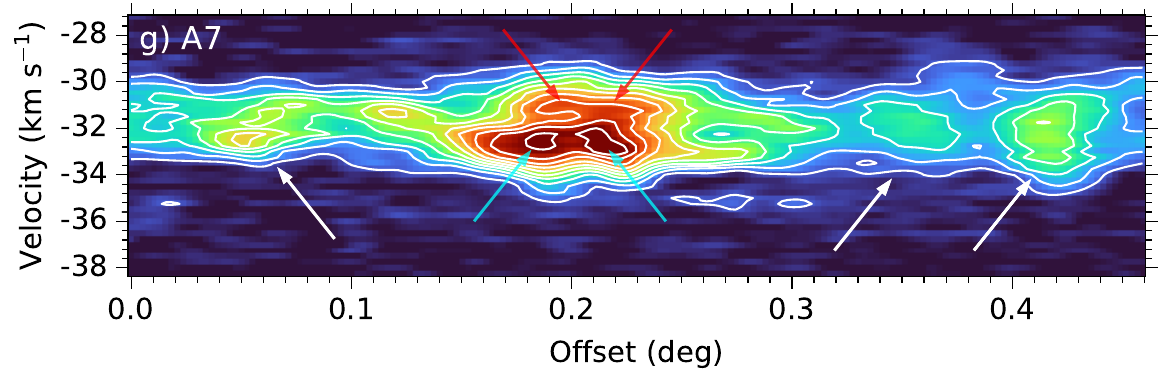}
\caption{Panels (a)--(g) present the {\it PV} diagrams along the arrows A1--A7, as indicated in Figure~\ref{321_twocolor}b. The contour values range from 1 to 10 K, with intervals of 1 K. The red and cyan arrows in each panel indicate two separate velocity components mixed together. The white arrows for A7 point to a single velocity component.}
\label{321_PV_diagrams}
\end{figure*}

\begin{figure*}
\centering
\includegraphics[width= 0.7\textwidth]{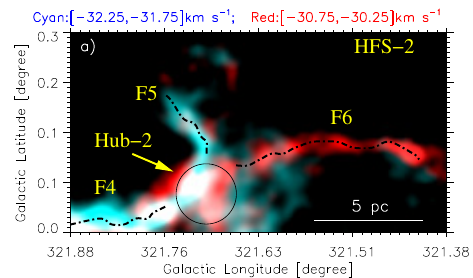}
\includegraphics[width= 0.7\textwidth]{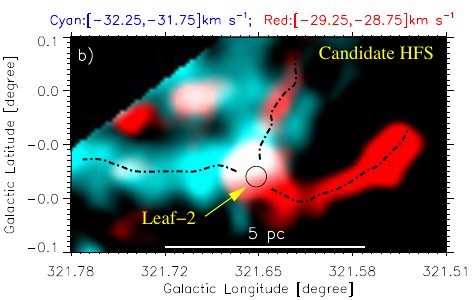}
\caption{(a) The two-color composite image presents the {\tco} integrated intensity maps shown in Figures~\ref{321_channel_maps}i (in cyan) and \ref{321_channel_maps}k (in red) for the region highlighted by the dashed-dotted rectangle in Figure~\ref{321_filament_maps}c. The moment-0 maps in cyan and red are displayed on a linear scale from 3$\sigma$ to 18$\sigma$ and 3$\sigma$ to 10$\sigma$, respectively, where $\sigma$ {\simi}0.15 K {\kps}. The black circle, indicated by a yellow arrow, is Hub-2 as shown in Figure~\ref{321_filament_maps}c. The dashed-dotted lines trace the filaments (i.e., F4, F5, and F6) connected to Hub-2.
(b) Similar two-color composite image showing {\tco} integrated intensity maps displayed in Figures~\ref{321_channel_maps}i (in cyan) and \ref{321_channel_maps}m (in red) for the area covered by the dashed-dotted polygon in Figure~\ref{321_ATLASGAL_154}a.
The moment-0 maps in cyan and red are displayed on a linear scale from 6$\sigma$ to 15$\sigma$ and 3$\sigma$ to 7$\sigma$, respectively. The black circle, situated at the peak emission of ATL-2 (indicated by a yellow arrow), has a radius equal to the leaf's effective radius. The dashed-dotted lines indicate the filaments connected to Leaf-2. A scale bar of 5 pc is shown in each panel.}
\label{2color_hub2_leaf2}
\end{figure*}

\section*{Acknowledgments}
We thank the anonymous referee for providing the valuable comments and suggestions, that improved the scientific content of this paper.
The research work at the Physical Research Laboratory is funded by the Department of Space, Government of India. We also acknowledge the support of Ajman University, Internal Research Grant No: [DRGS Ref. 2024-IRG-HBS-7]. A.K.M. thanks A. Men’shchikov for scientific discussions on {\it hires}.
This work is based [in part] on observations made with the {\it Spitzer} Space Telescope, which is operated by the Jet Propulsion Laboratory, California Institute of Technology, under a contract with NASA. This research has made use of the NASA/IPAC Infrared Science Archive, which is funded by the National Aeronautics and Space Administration and operated by the California Institute of Technology. These observations are associated with the proposal ID 14465. This paper makes use of the following ALMA data: ADS/JAO.ALMA\#2013.1.00960.S. ALMA is a partnership of ESO (representing its member states), NSF (USA) and NINS (Japan), together with NRC (Canada), MOST and ASIAA (Taiwan), and KASI (Republic of Korea), in cooperation with the Republic of Chile. The Joint ALMA Observatory is operated by ESO, AUI/NRAO and NAOJ. The MeerKAT telescope is operated by the South African Radio Astronomy Observatory, which is a facility of the National Research Foundation, an agency of the Department of Science and Innovation. This research has made use of the VizieR catalogue access tool, CDS, Strasbourg, France (DOI : 10.26093/cds/vizier). The original description of the VizieR service was published in 2000, A\&AS 143, 23. This research made use of {\it Astropy}\footnote[1]{http://www.astropy.org}, a community-developed core Python package for Astronomy \citep{astropy13,astropy18}. For figures, we have used {\it matplotlib} \citep{Hunter_2007} and IDL software.  
%

\bibliographystyle{aasjournal}
\bibliography{reference}{}
\appendix
\restartappendixnumbering

\section{The optical depth and column density maps} \label{sec_appendix_tau13}
\begin{figure*}[bt]
\centering
\includegraphics[width= 0.9\textwidth]{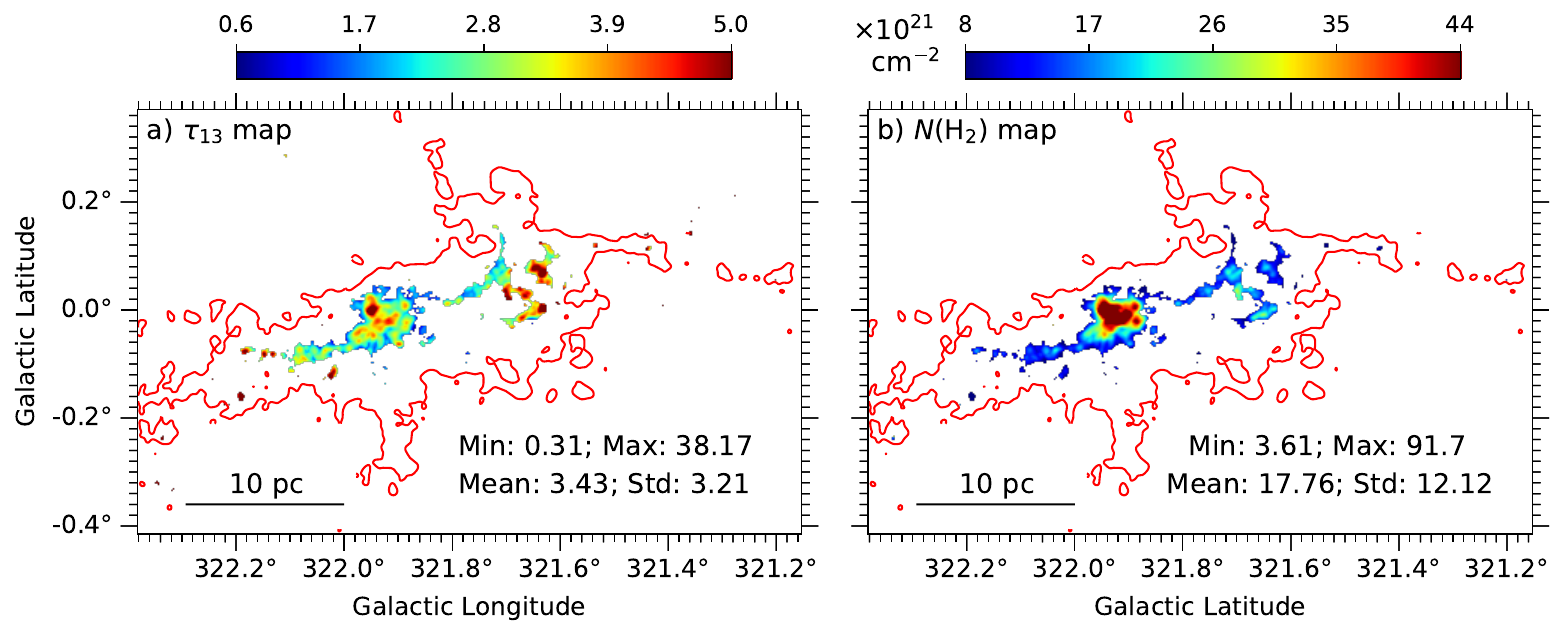}
\caption{(a) The optical depth (i.e., $\tau_{13}$) map. The minimum (Min), maximum (Max), mean, and standard deviation (Std) values for the $\tau_{13}$ map are provided in the panel. (b) The column density (i.e., $N$(H$_2$)) map for $T{_\mathrm{ex}} = 15$ K. The Min, Max, mean, and Std values for the $N$(H$_2$) map are indicated in the panel in units of $10^{21}$ cm$^{-2}$. In both panels, the red contour indicates the extent of the {\tco} emission above the 3$\sigma$ limit, as shown in Figure~\ref{321_molecular_maps}. A scale bar of 10 pc is displayed in each panel.}
\label{tau13andNH2}
\end{figure*}

\section{The estimation of the dynamical age of the {\htwo} regions} \label{sec_appendix_sptypeHII}
The dynamical age of the {\htwo} regions, which are assumed to be expanding in a uniform medium, can be estimated using the formula from \citet{Dyson_1980}:
\begin{equation}
{t_{\rm  dyn}}=\frac{4R_{\rm s}}{7c_{\rm s}}\left[\left(\frac{R_{\rm H {\small II}}}{R_{\rm s}}\right)^{7/4}-1\right],
\label{tdyn}
\end{equation}
where $c_{\rm s}$ \citep[{\simi}10 km s$^{-1}$;][]{Bisbas_2009} is the sound speed in the ionized region. $R_{\rm H {\small II}}$ is the effective radius of the {\htwo} region, which was calculated based on their area above 5$\sigma$ in the MeerKAT 1.28 GHz radio continuum data (see Figure~\ref{G321_herschel_maps}c). The $R_{\rm H {\small II}}$ value for each {\htwo} region is listed in Table~\ref{tab1_htwo}.
$R_{\rm s}$ is the Str\"{o}mgren radius, given by 
$R_{\rm s} = \left( {3N_{\rm uv}}/{4\pi n^{2}_{\rm i} \alpha_{\rm B}} \right)^{1/3}$, where $N_{\rm uv}$ is the number of photons emitted by the ionizing source beyond the Lyman limit, $n_{\rm i}$ is the initial H number density, and $\alpha_{\rm B}$ \citep[{\simi}2.6$\times10^{-13}$ cm$^{3}$ s$^{-1}$;][]{Kwan_1997} is the recombination coefficient. We used $n_{\rm i}$ = 10$^4$ cm$^{-3}$ based on the average H$_{2}$ density for the ATLASGAL leaves (see Table~\ref{tab_atlasgal_tree}). We calculated $N_{\rm uv}$ using the formula \citep{Matsakis_1976AJ}:
\begin{equation}
 {N_{\rm uv}} ~[s^{-1}]=7.54\times10^{46}\,\left(\frac{\nu}{\rm GHz}\right)^{0.1}\left(\frac{T_{\rm {eff}}}{10^4\,\rm K}\right)^{-0.45}\left(\frac{S_{\nu}}{\rm Jy}\right)\,\left(\frac{d}{\rm kpc}\right)^{2}\rm ,
\end{equation}
where, $\nu = 1.28$ GHz, $d = 1.98$ kpc, $T_{\rm eff}$ is the effective electron temperature (assumed to be 10$^{4}$ K), $S_{\nu}$ is the total flux density for the {\htwo} regions. Our calculated $S_{\nu}$ (above $>5\sigma$) and corresponding log($N_{\rm uv} ~[s^{-1}]$) values are included in Table~\ref{tab1_htwo}. Finally, using all the input parameters in Equation~\ref{tdyn}, we estimated the dynamical time scale for the {\htwo } regions and listed in Table~\ref{tab1_htwo}.

\end{document}